\documentclass[10pt,pdftex]{iopart}
\usepackage{iopams}

\usepackage{tensind,graphicx,gensymb,bbm,wasysym} %,amssymb,gensymb,bbm}
\usepackage{mathptmx}      % use Times fonts if available on your TeX system
\usepackage{color}

\def\sech{\textrm{sech}}
\def\arsinh{\textrm{arsinh}}

\begin{document}
\tensordelimiter{?}

\title{Detailed study of null and time-like geodesics in the Alcubierre Warp spacetime}

\author{Thomas M{\"u}ller}
\address{
  Visualisierungsinstitut der Universit\"at Stuttgart (VISUS)\\
  Allmandring 19, 70569 Stuttgart, Germany
}
\ead{Thomas.Mueller@visus.uni-stuttgart.de}

\author{Daniel Weiskopf}
\address{
  Visualisierungsinstitut der Universit\"at Stuttgart (VISUS)\\
  Allmandring 19, 70569 Stuttgart, Germany
}
\ead{Daniel.Weiskopf@visus.uni-stuttgart.de}

% -----------------------------------------------------------------
%                            Abstract
% -----------------------------------------------------------------
\begin{abstract}
  The Alcubierre warp spacetime yields a fascinating chance for comfortable interstellar travel between arbitrary distant places without the time dilation effect as in special relativistic flights. Even though the warp spacetime needs exotic matter for its construction and is thus far from being physically feasible, it offers a rich playground for studying geodesics in the general theory of relativity.   
  This paper is addressed to graduate students who have finished a first course in general relativity to give them a deeper inside in the calculation of non-affinely parametrized null and time-like geodesics and a straightforward approach to determine the gravitational lensing effect due to curved spacetime by means of the Jacobi equation. Both topics are necessary for a thorough discussion of the visual effects as observed by a traveller inside the warp bubble or a person looking from outside. The visual effects of the traveller can be reproduced with an interactive Java application.
\end{abstract}

% -----------------------------------------------------------------
%                            PACS
%
%   http://publish.aps.org/PACS
% -----------------------------------------------------------------

%% 04.20.-q   Classical general relativity
%% 04.25.Dm   Numerical relativity
%% 89.20.Ff   Computer science and technology
%% 95.75.Pq   Mathematical procedures and computer techniques

\submitto{\EJP}

%\keywords{Alcubierre Warp spacetime, null and timelike geodesics, interactive visualization}
%\maketitle

% -----------------------------------------------------------------
%                            Introduction
% -----------------------------------------------------------------
\section{Introduction}\label{sec:intro}
Even with the very latest propulsion technology we are far from being able to comfortably travel around in our own solar system, not to mention to travel to extra-solar planets. But even if one day we would have the necessary rocket drives and we could accelerate to nearly the speed of light to overcome the enormous distances, the time dilation effect of special relativity would be boon and bane together. In a certain way, general relativity could be a way out because it offers several mathematical solutions of the Einstein field equations like the Morris--Thorne wormhole~\cite{morris} or the warp metric by Alcubierre~\cite{alcubierre1994} which could be used for interstellar travel. While a wormhole represents a shortcut in spacetime, the warp metric contracts and expands the spacetime locally and encompasses like a bubble a nearly flat region that is, in some sense, decoupled from the rest of the spacetime. Unfortunately, both mathematical solutions need exotic matter, which makes them physically unrealizable.

Nonetheless, these spacetimes offer a rich playground for studying geodesics in general relativity. While geodesics in the Morris--Thorne wormhole can be handled numerically in a quite straightforward manner, the time-dependent Alcubierre spacetime makes it necessary to transform the geodesic equation, the equation for the parallel transport of vectors along time-like geodesics, and the equations to determine the gravitational lensing effect caused by the warp bubble into their non-affinely parametrized form. Then, the resulting equations can be integrated numerically as usual.

The aim of our article is to study in detail the influence of the warp bubble on null and time-like geodesics. For that, we first transform all the relevant equations mentioned above into their non-affinely parametrized form. Beside the paths of light rays, we also discuss the frequency shift and the effect on a bundle of light rays resulting in gravitational lensing that let us visualize point-like objects in a most realistic way. By means of the parallel-transport equation, we show how particles initially at rest will be carried along and undergo geodesic precession. Furthermore, we explain what an observer would actually see inside or outside the warp bubble using either four-dimensional ray tracing or an interactive Java application. For the latter, we make use of the highly efficient hardware architecture of programmable graphical processing units (GPUs) to correctly visualize point-like stars in contrast to the elongated stars shown in science fiction movies. The view from inside the warp bubble will be compared to the view of a special relativistic observer. The differences between the views of these observers can be clearly reproduced using our Java application.

A detailed discussion of null geodesics from inside the warp bubble, in particular the emergence of an apparent horizon behind the warp bubble, was given by Clark et al.~\cite{clark1999} What an observer within the Alcubierre spacetime would actually see was shown by Weiskopf~\cite{weiskopfDiss}. There are several papers that discuss the physical nature of warp drive spacetimes, and we can give only a few references.\cite{hiscock1997,pfenning1997,broeck1999,lobo2004} 

To study geodesics in detail, we refer the reader to the interactive visualization tool {\it GeodesicViewer}~\cite{mueller2010a,mueller2011a} and how it could be used in the classroom.

The structure of this paper is as follows. In section \ref{sec:warp}, we briefly review the Alcubierre warp metric and present the local reference frames of a comoving and a static observer. In section \ref{sec:nullGeodesics}, we study the trajectories of light rays and how they influence the view of an observer that either co-moves with the warp bubble or stays static in the outside. We also discuss the frequency shift and the lensing effect caused by the warp bubble. For the case of a comoving observer, we present our interactive Java application and the necessary technical and implementation details in section \ref{sec:vis}. Finally, we discuss the influence of the warp bubble on particle trajectories in section \ref{sec:timelikeGeodesics}. The technical details of the integration of the non-affinely parametrized equations are relegated to the appendix.

The Java application and its sources, high-resolution images of this article, as well as some movies that show the visual distortion of the warp bubble or the motion of particles can be downloaded from 
%\url{http://www.vis.uni-stuttgart.de/relativity}.}
\textcolor{magenta}{ http://www.vis.uni-stuttgart.de/\~{}muelleta/Warp }

% -----------------------------------------------------------------
%                    Warp metric revisited
% -----------------------------------------------------------------
\section{Warp metric and local reference frames}\label{sec:warp}
\begin{figure}[ht]
 \centering
 \includegraphics[scale=0.72]{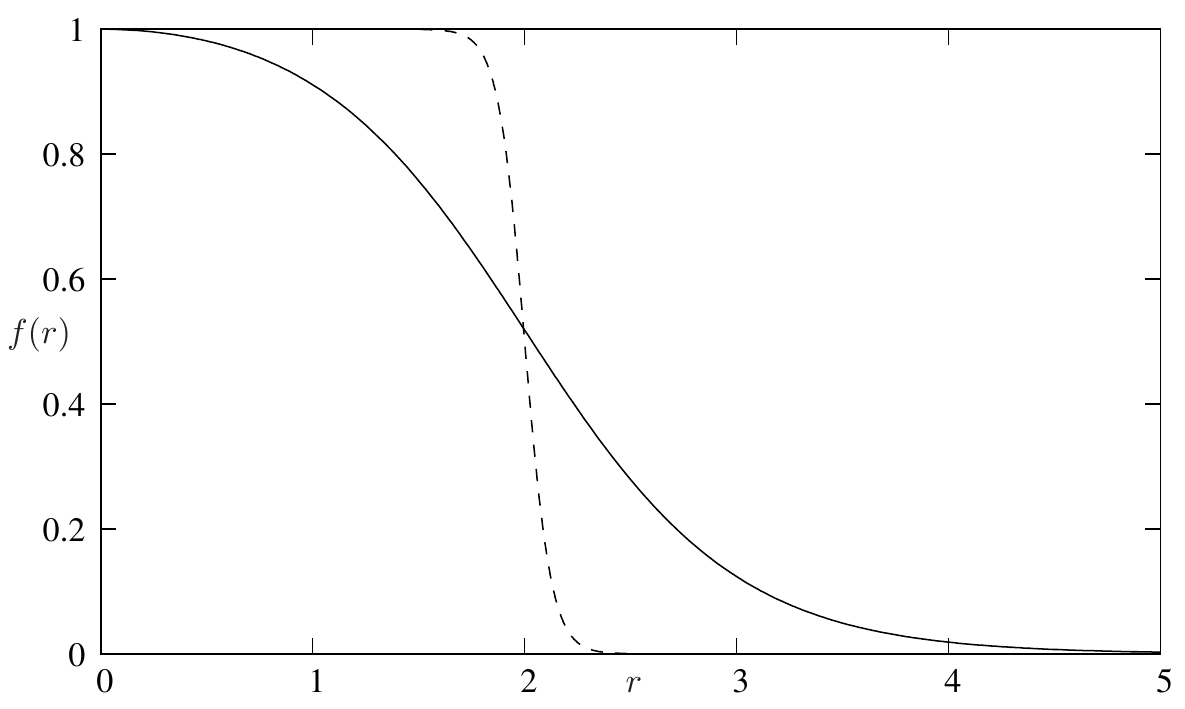}
 \caption{Shape function $f(r)$ for a radius $R=2$ and thickness parameters $\sigma=1$ (solid line) and $\sigma=8$ (dashed line).}
 \label{fig:shapefunc}
\end{figure}

The Warp metric developed by Miguel Alcubierre~\cite{alcubierre1994} can be described by the line element
\begin{equation}
  \label{eq:warpMetric}
  ds^2 = -c^2dt^2+\left[dx-vf(r)dt\right]^2 + dy^2 + dz^2,
\end{equation}
where $c$ is the speed of light,
\begin{eqnarray}
  v  &= \frac{dx(t)}{dt},\\
  r(t) &= \sqrt{(x-x(t))^2+y^2+z^2},\\
  \label{eq:formfunc} f(r) &= \frac{\tanh (\sigma (r+R)) - \tanh (\sigma (r-R))}{2\tanh (\sigma R)},
\end{eqnarray}
and $x(t)$ is the worldline of the center of the warp bubble. The parameters $R>0$ and $\sigma>0$  in the shape function $f(r)$ define the radius and the thickness of the bubble, see figure \ref{fig:shapefunc}.

For this metric, we can define two natural local tetrads $\mathbf{e}_{(i)}=e_{(i)}^{\mu}\partial_{\mu}$ that represent the local reference frames of either a comoving or a static observer. 
The comoving tetrad is defined by
\begin{equation}
 \label{eq:comLT}
 \mathbf{e}_{(0)} = \frac{1}{c}\left(\partial_t + vf\partial_x\right), \,
 \mathbf{e}_{(1)} = \partial_x,\,
 \mathbf{e}_{(2)} = \partial_y,\,
 \mathbf{e}_{(3)} = \partial_z,
\end{equation}
% with its dual tetrad $\mathbf{\theta}^{(i)}=\theta^{(i)}_{\mu}dx^{\mu}$ given by
% \begin{equation}
%  \mathbf{\theta}^{(0)} = c\,dt,\quad
%  \mathbf{\theta}^{(1)} = -vf\,dt + dx,\quad
%  \mathbf{\theta}^{(2)} = dy,\quad
%  \mathbf{\theta}^{(3)} = dz.
% \end{equation}
% The local tetrad and its dual are related via $\mathbf{\theta}^{(i)}\mathbf{e}_{(j)}=\delta_{(j)}^{(i)}$. The static local tetrad reads
and the static local tetrad reads
\begin{eqnarray}
 \mathbf{\hat{e}}_{(0)} &= \frac{1}{\sqrt{c^2-v^2f^2}}\partial_t,\quad \mathbf{\hat{e}}_{(2)} = \partial_y,\quad \mathbf{\hat{e}}_{(3)} = \partial_z,\\
 \mathbf{\hat{e}}_{(1)} &= -\frac{vf}{c\sqrt{c^2-v^2f^2}}\partial_t + \frac{\sqrt{c^2-v^2f^2}}{c}\partial_x.
\end{eqnarray}
It is obvious that the comoving tetrad is valid everywhere, whereas the static tetrad is defined only in the region of the spacetime where $v^2f^2<c^2$. Both tetrads fulfill the orthonormality condition $g_{\mu\nu}e_{(i)}^{\mu}e_{(j)}^{\nu}=\eta_{(i)(j)}$ with $\eta_{(i)(j)}=\mbox{diag}(-1,1,1,1)$, which means that these tetrads locally define a Minkowskian system.

Throughout the paper we consider a warp bubble that moves with constant velocity. Thus, the center of the warp bubble follows the worldline $x(t)=vt$. Furthermore, we set $\{c\}=1$ for numerical examples. Then, times and distances are given in seconds and light-seconds or years and light-years, respectively.

% -----------------------------------------------------------------
%                      null geodesics
% -----------------------------------------------------------------
\section{Null geodesics}\label{sec:nullGeodesics}
In this section, we will discuss the influence of the warp bubble on the propagation of light for several different situations. In general, light paths follow from the numerical integration of the null geodesic equation 
\begin{equation}
 \frac{d^2x^{\mu}}{d\lambda^2}+\Gamma_{\nu\rho}^{\mu}\frac{dx^{\nu}}{d\lambda}\frac{dx^{\rho}}{d\lambda}=0,
\end{equation}
where $\lambda$ is an affine parameter and $\Gamma_{\nu\rho}^{\mu}$ are the Christoffel symbols of the second kind (see App.~\ref{app:details} for the Christoffel symbols of the Warp metric). After each integration step, we have to check that the constraint equation 
\begin{equation}
  g_{\mu\nu}\frac{dx^{\mu}}{d\lambda} \frac{dx^{\nu}}{d\lambda} = 0
  \label{eq:constrEq}
\end{equation}
with metric tensor $g_{\mu\nu}$ is still fulfilled. This constraint ensures that the geodesic remains light-like even if numerical error is present. 

For the warp metric, an integrator with step-size control is indispensable. Otherwise, the step-size would have to be inefficiently tiny. 
However, irrespective of the numerical integrator, direct integration of the geodesic equation leads to numerical problems at the rim of the bubble for certain initial values because of the inappropriate affine parameter. Although the spacetime coordinates $(t,x,y,z)$ are smooth, the step-size of the affine parameter becomes extremely small. Thus, an integrator with step-size control will get stuck and the constraint equation will be violated.
To avoid the numerical difficulties, we use the non-affinely parametrized geodesic as described in App.~\ref{app:naGeod}. Additionally, the resulting equations are numerically much more accurate than the ones that follow from the affinely parametrized geodesic equation.

The gravitational frequency shift $z_f$ between the emitted, $\omega_{\mbox{src}}$, and the observed, $\omega_{\mbox{obs}}$, light frequencies is obtained by
\begin{equation}
 1+z_f = \frac{\omega_{\mbox{src}}}{\omega_{\mbox{obs}}} = \frac{g_{\mu\nu}u_{\mbox{src}}^{\mu}k^{\nu}}{g_{\mu\nu}u_{\mbox{obs}}^{\mu}k^{\nu}},
\end{equation}
where $\mathbf{u}_{\mbox{src}}$ is the four-velocity of the light source and $\mathbf{u}_{\mbox{obs}}$ is the four-velocity of the observer, see e.g. Wald~\cite{wald}. The tangent of the light ray is given by $k^{\mu}=dx^{\mu}/d\lambda$ and must be evaluated either at the source or the observer position. If $-1<z_f<0$, we call it a blueshift, and if $z_f>0$, it is a redshift.

To determine the lensing effect caused by the warp bubble, we study the behavior of the spacetime curvature on a bundle of light rays that is described by two Jacobian fields $\mathbf{Y}_i=Y_i^{\mu}\partial_{\mu}$. The change of the Jacobian fields along the central light ray with tangent $\mathbf{k}=k^{\mu}\partial_{\mu}$ is determined by the Jacobian equation
\begin{equation}
 \frac{D^2Y_i^{\mu}}{d\lambda^2} = ?R^{\mu}_{\nu\rho\sigma}?k^{\nu}k^{\rho}Y_i^{\sigma}.
\end{equation}
The cross section of the light bundle follows from the projection of the Jacobian fields onto the parallel-transported Sachs vectors $\mathbf{s}_i=s_i^{\mu}\partial_{\mu}$ that are perpendicular to the light ray $\mathbf{k}$. The resulting Jacobi matrix
\begin{equation}
 J_{ij}=g_{\mu\nu}Y_i^{\mu}s_j^{\nu}
\end{equation}
describes how the shape of the initially circular bundle of light rays transforms into an ellipse with major and minor axes $a_{\pm}$ along the central light ray. Fortunately, we can calculate the matrix by first integrating from the observer to the source and then inverting this matrix to obtain the behavior of a light bundle from the source to the observer. The result of this calculation is the magnification factor $\mu_{\mbox{mag}}=\lambda^2/(a_{+}a_{-})$. 
%\textcolor{red}{In brief, the magnification factor gives the ratio $d\Omega_{\mbox{lensed}}/d\Omega_{\mbox{unlensed}}$ between the lensed and the unlensed solid angle of an object.} 
Details of this calculation in the context of the non-affinely parametrized geodesic equation can be found in App.~\ref{app:sachsjac}. A thorough discussion of gravitational lensing, however, is out of the scope of this article and we refer the interested reader to the standard literature~\cite{schneider1992} (See also Ohanian~\cite{ohanian} for a discussion of the gravitational lensing by a Schwarzschild black hole.)

%% ---------------------------------------------------------------------
\subsection{View from the bridge}
The view from the bridge was already discussed in detail by Clark et al.~\cite{clark1999} Here, we will reconstruct their results for the sake of completeness. Additionally, we will show how light rays behave in the close neighborhood of the warp bubble and how the warp bubble influences the lensing of point-like objects.

The local reference frame of the bridge is given by the comoving tetrad of Eq.~(\ref{eq:comLT}). Because of the axial symmetry, we can restrict to geodesics in the $xy$-plane. Then, an incident light ray with angle $\xi$ with respect to the local reference frame can be described by the four vector
\begin{equation}
 \mathbf{k}=\omega\left(-\mathbf{e}_{(0)} + \cos\xi\mathbf{e}_{(1)} + \sin\xi\mathbf{e}_{(2)}\right) = k^{\mu}\partial_{\mu}.
 \label{eq:initK}
\end{equation}
Here, we use the minus sign in front of the time-like tetrad vector $\mathbf{e}_{(0)}$ because we integrate geodesics back in time. If we are only interested in the paths of the light rays, we can set the frequency $\omega=1$.

\begin{figure}[ht]
 \centering
 \includegraphics[scale=0.8]{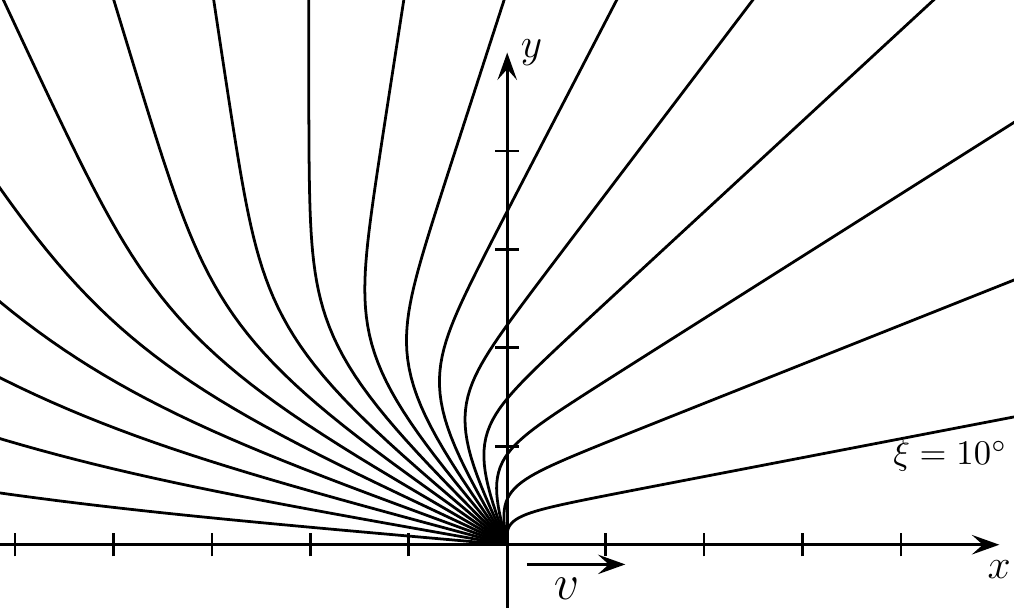}
 \caption{Light rays reach the comoving observer on the bridge, here at $(t=0,x=0,y=0)$, with incident angles $\xi=\left\{10\degree,20\degree,\ldots,170\degree\right\}$ with respect to the comoving local reference frame $\mathbf{e}_{(i)}$. The parameters of the warp metric read $R=2$, $\sigma=1$, $v=c$.}
 \label{fig:bridgeIncident}
\end{figure}

Figure~\ref{fig:bridgeIncident} shows several light rays for an observer on the bridge when the observer passes the origin at $t=0$. Although all the incident light rays are equidistantly separated by $10\degree$ with respect to the observer's local reference frame, they apparently approach the observer from behind.

\begin{figure}[ht]
 \centering
 \includegraphics[scale=0.3]{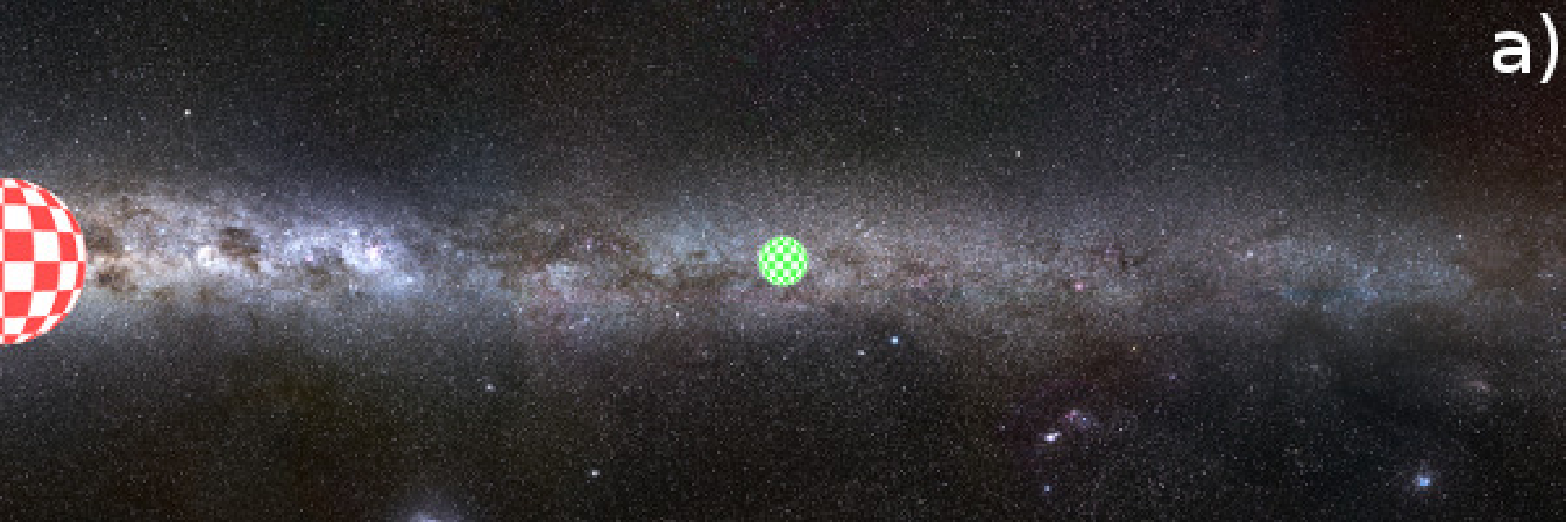}\\
 \includegraphics[scale=0.3]{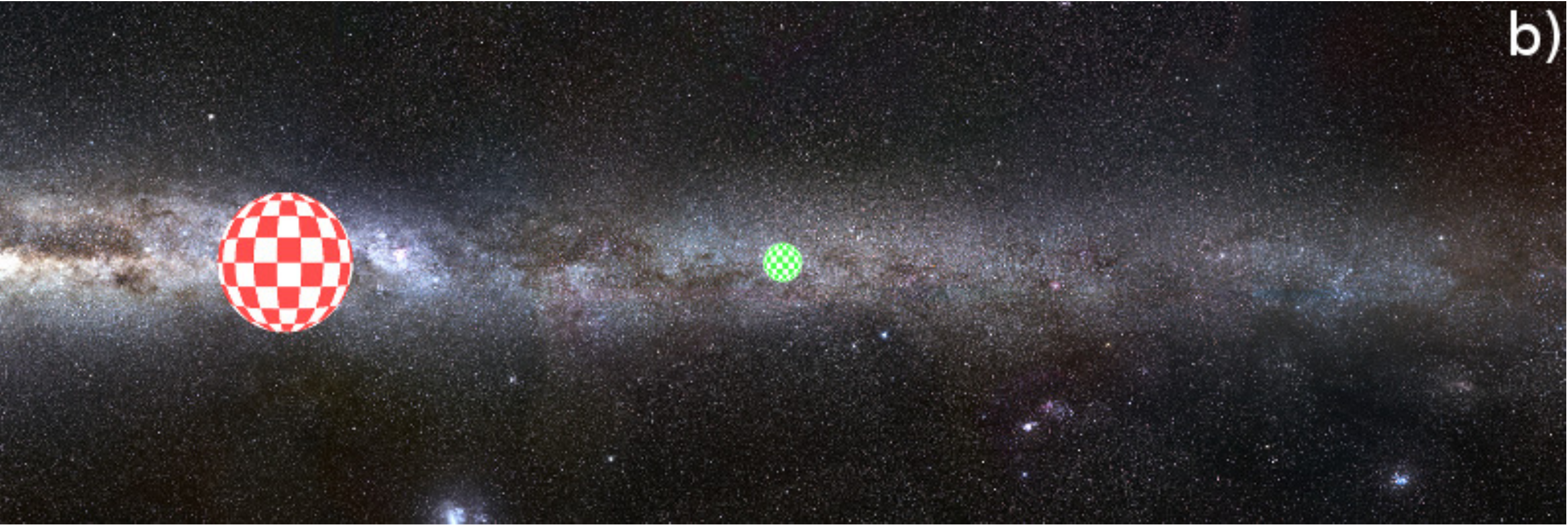}\\
 \includegraphics[scale=0.3]{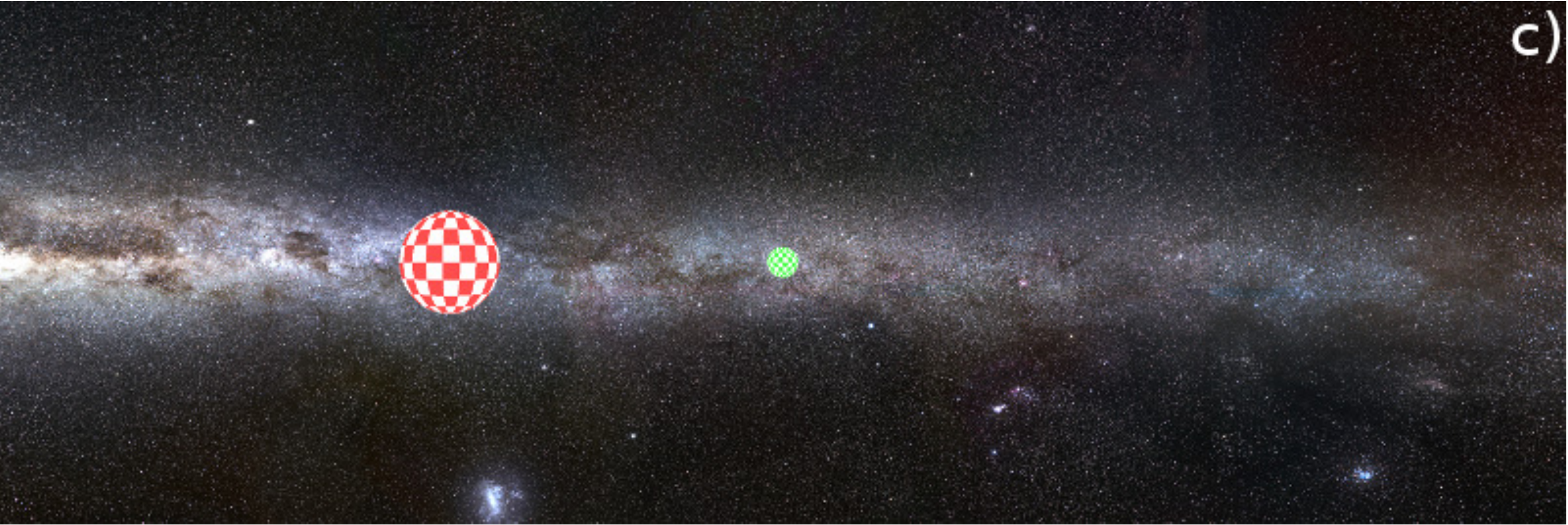}
 \caption{View from the bridge at $t=0$ in the direction of motion for velocities $v=0.01c$ (a), $v=c$ (b), $v=2c$ (c) and a panorama camera with $180\degree\times 60\degree$ field of view. The Milky Way background is represented by a sphere with radius $r_{\mbox{max}}=200$.\cite{eso}}
 \label{fig:bridgeFP}
\end{figure}

Figure~\ref{fig:bridgeFP} shows the view from the bridge of an observer moving with different warp speeds when passing the origin $x=y=0$ at $t=0$. To visualize distortion effects close to the warp bubble, we use two checkered balls of radius $r_{\mbox{ball}}=0.5$. The green ball is located at $x=10, y=0$ whereas the red ball is located at $x=0, y=3$. Similar to the special relativistic motion in flat Minkowskian spacetime, there is an aberration in the direction of motion that is stronger the faster the warp bubble moves. In contrast to the special relativistic motion, however, the aberration here is only due to the curved spacetime.

\begin{figure}[ht]
 \centering
 \includegraphics[scale=0.72]{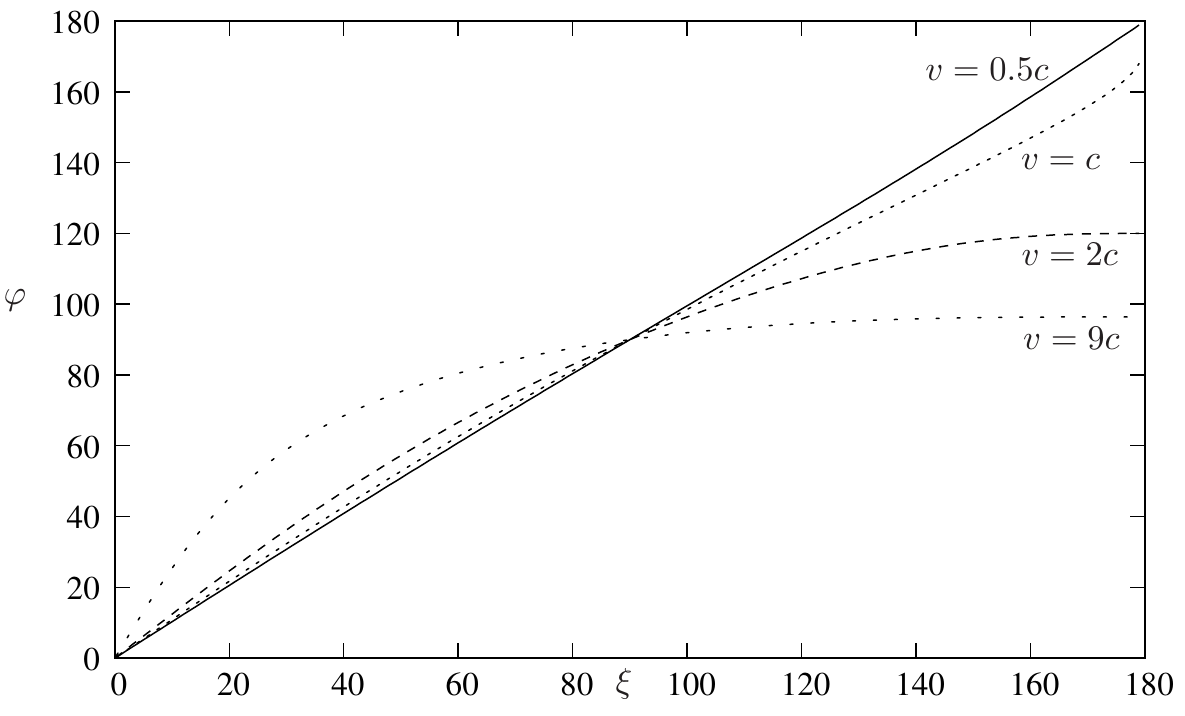}
 \caption{Angle $\varphi$ at $r_{\mbox{max}}=5\times 10^4$ with respect to the incident angle $\xi$ for velocities $v/c=\left\{0.5,1,2,9\right\}$ and warp parameters $R=2$, $\sigma=1$. The angle $\xi=0\degree$ corresponds to the direction of motion.}
 \label{fig:bridgeInf}
\end{figure}

As already found by Clark et al. there is no aberration for $\xi=90\degree$. Hence, light rays that originate precisely at $90\degree$ to the direction of motion of the warp bubble will be seen by an observer at the bridge at an angle $\xi=90\degree$ (see also App.~\ref{app:euler}). Furthermore, the observer on the bridge cannot see the whole spacetime. If we trace light rays from the bridge with initial angles $\xi$ back in time until they hit a sphere in the asymptotic background, $r=r_{\mbox{max}}$, cf. figure \ref{fig:bridgeInf}, we find that there is an apparent horizon opposite to the direction of motion, where the size of the apparent horizon depends on the velocity of the warp bubble. However, the observer will not see a black region because for any direction $\xi$ some region of the asymptotic background will be visible.

\begin{figure}[ht]
 \centering
 \includegraphics[scale=0.72]{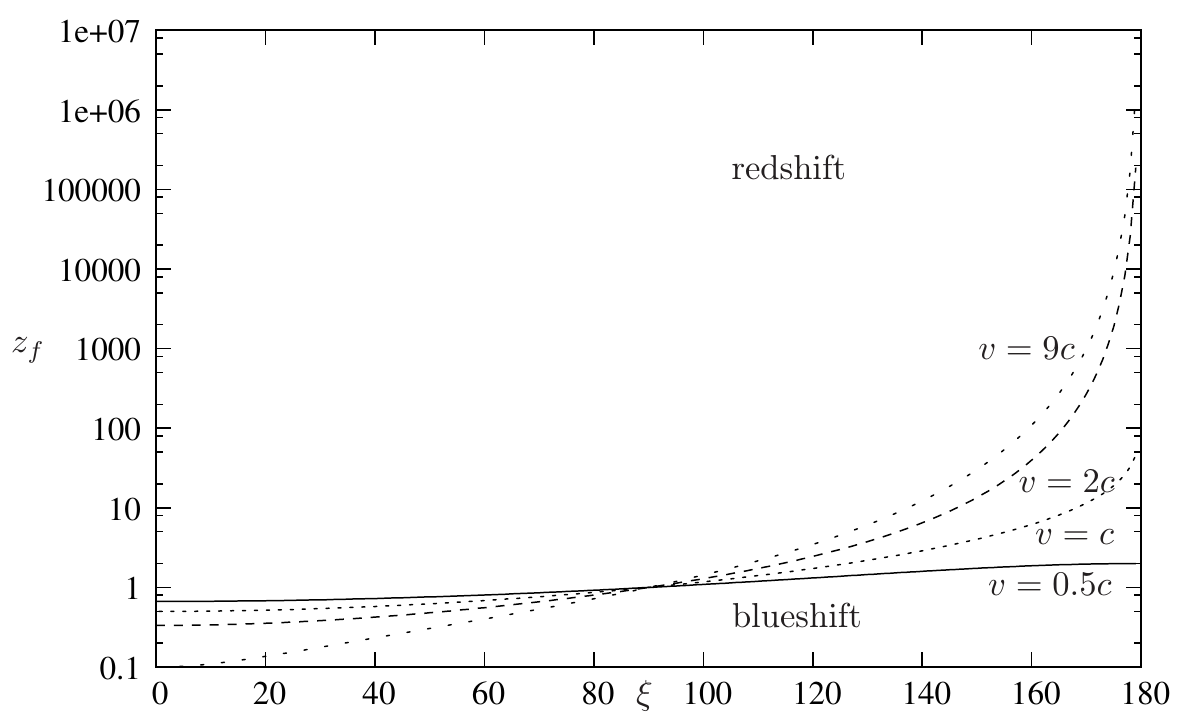}
 \caption{Frequency shift $1+z_f$ between $r_{\mbox{max}}=5\times 10^4$ and the bridge with respect to the incident angle $\xi$ for velocities $v/c=\left\{0.5,1,2,9\right\}$ and warp parameters $R=2$, $\sigma=1$.}
 \label{fig:freqBridge}
\end{figure}

To determine the frequency shift and the lensing caused by the warp bubble, we use the asymptotic sphere as the place where there are everywhere point-like light sources with unit frequency $\omega_{\mbox{src}}=1$. Then, the observer at the bridge will detect frequency shifts as shown in figure \ref{fig:freqBridge}. Irrespective of the velocity, there is no frequency shift for $\xi=90\degree$. In the direction of motion, $\xi<90\degree$, we have a slight blueshift, whereas in the opposite direction the redshift is quite strong.

\begin{figure}[ht]
 \centering
 \includegraphics[scale=0.72]{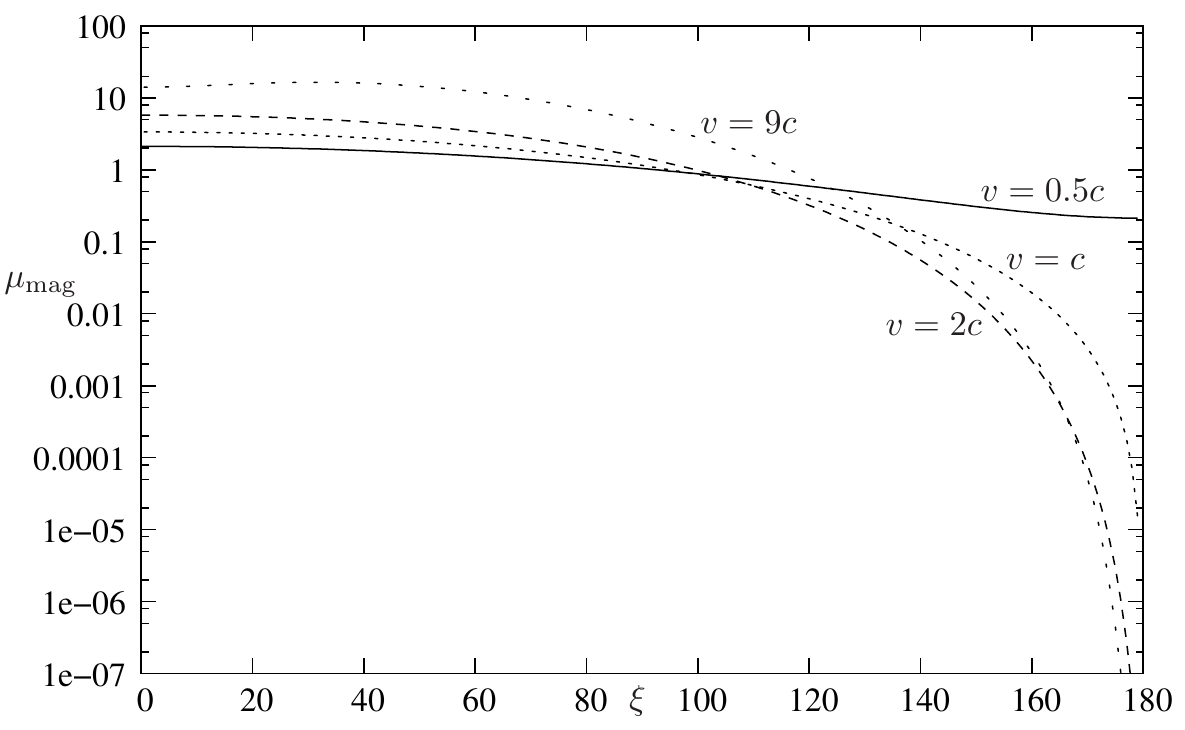}
 \caption{Lensing $\mu_{\mbox{mag}}$ between $r_{\mbox{max}}=5\times 10^4$ and the bridge with respect to the incident angle $\xi$ for velocities $v/c=\left\{0.5,1,2,9\right\}$ and warp parameters $R=2$, $\sigma=1$.}
 \label{fig:magBridge}
\end{figure}

The lensing of point-like sources caused by the warp bubble is shown in figure \ref{fig:magBridge}. In the direction of motion, we have a slight magnification, whereas in the opposite direction, the light of point-like sources are strongly dimmed. 

Hence, even though the horizon does not produce a black region, the strong redshift together with the magnification let the region $\xi<90\degree$ appear dark for high velocities. 

The `view from the bridge' for a special-relativistically moving observer is well-known from standard literature. The aberration and Doppler-shift can be derived from the representation of an initial direction $\mathbf{k}$ with respect to either the moving reference frame
\begin{equation}
 \mathbf{\bar{e}}_{(0)} = \frac{\gamma}{c}\left(\partial_t+v_{\mbox{sr}}\partial_x\right),\quad\mathbf{\bar{e}}_{(1)}=\gamma\left(v_{\mbox{sr}}\frac{\partial_t}{c}+\partial_x\right),
\end{equation}
$\mathbf{\bar{e}}_{(2)}=\partial_y$, $\mathbf{\bar{e}}_{(3)}=\partial_z$, as in Eq.~(\ref{eq:initK}), or the standard Minkowskian frame. Here, $v_{\mbox{sr}}$ is the special-relativistic velocity of the moving observer. Thus, we obtain the aberration formula
\begin{equation}
  \cos\xi' = \frac{\gamma}{D}\left(\cos\bar{\xi}-v_{\mbox{sr}}/c\right)
\end{equation}
and the Doppler-shift
\begin{equation}
 1+z_f = D,
\end{equation}
where $D=\gamma[1-(v_{\mbox{sr}}/c)\cos\bar{\xi}]$ is the Doppler-factor and $\gamma=1/\sqrt{1-v_{\mbox{sr}}^2/c^2}$. Here, the primed angle is with respect to the flat Minkowski spacetime and the barred angle is with respect to the special-relativistic reference frame. The magnification factor is given by
\begin{equation}
 \mu_{\mbox{mag}} = D^{-2},
\end{equation}
see Weiskopf et al.~\cite{weiskopf99} for a detailed discussion. For velocities $v_{\mbox{sr}}$ close to the speed of light, we have very strong blueshift and huge magnification in the direction of motion. 

While the direction of no frequency shift, $z_f=0$, in the Warp metric is fixed by $\xi=90\degree$, this borderline is defined by $D=1$ in the special-relativistic case, which also gives the borderline of no-magnification, $\mu_{\mbox{mag}}=1$. The corresponding angle reads $\xi=\arccos\left[(\gamma-1)/(\gamma\beta)\right]$, where $\xi<90\degree$ for $v_{\mbox{sr}}>0$. Hence, for velocities close to the speed of light, a special-relativistic traveller will only see a small bright spot in contrast to a traveller inside a warp bubble.

%% ---------------------------------------------------------------------
\subsection{Warp bubble approaching a static observer}
\begin{figure}[ht]
 \centering
 \includegraphics[scale=0.3]{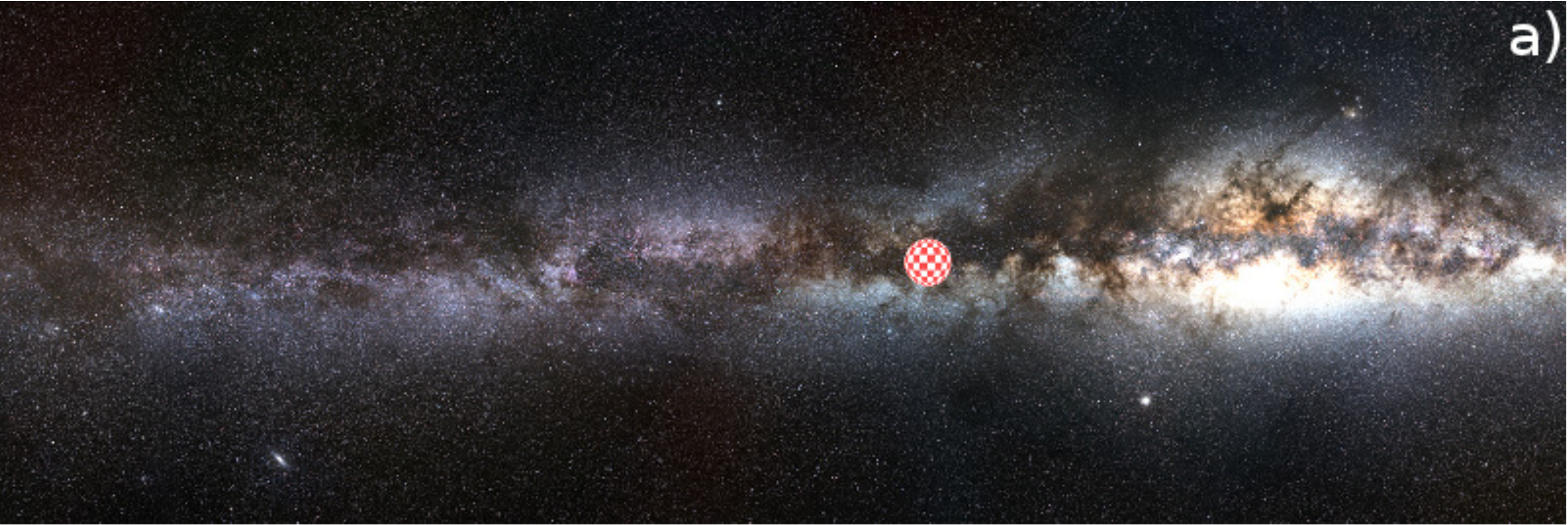}\\
 \includegraphics[scale=0.3]{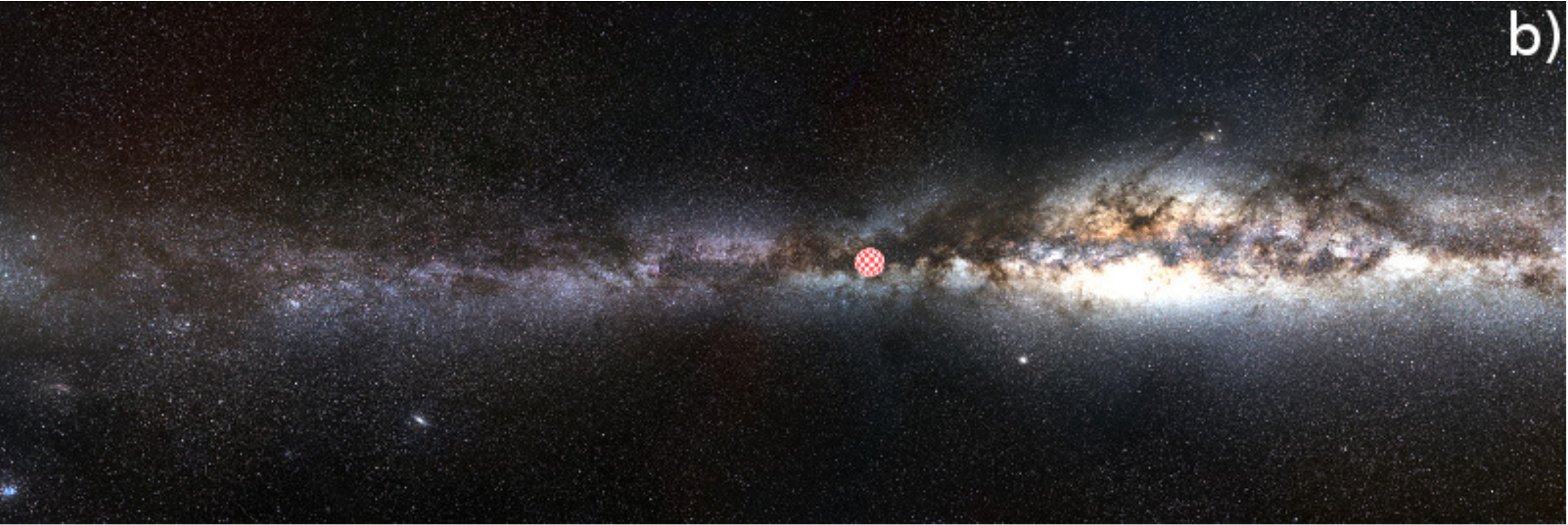}\\
 \includegraphics[scale=0.3]{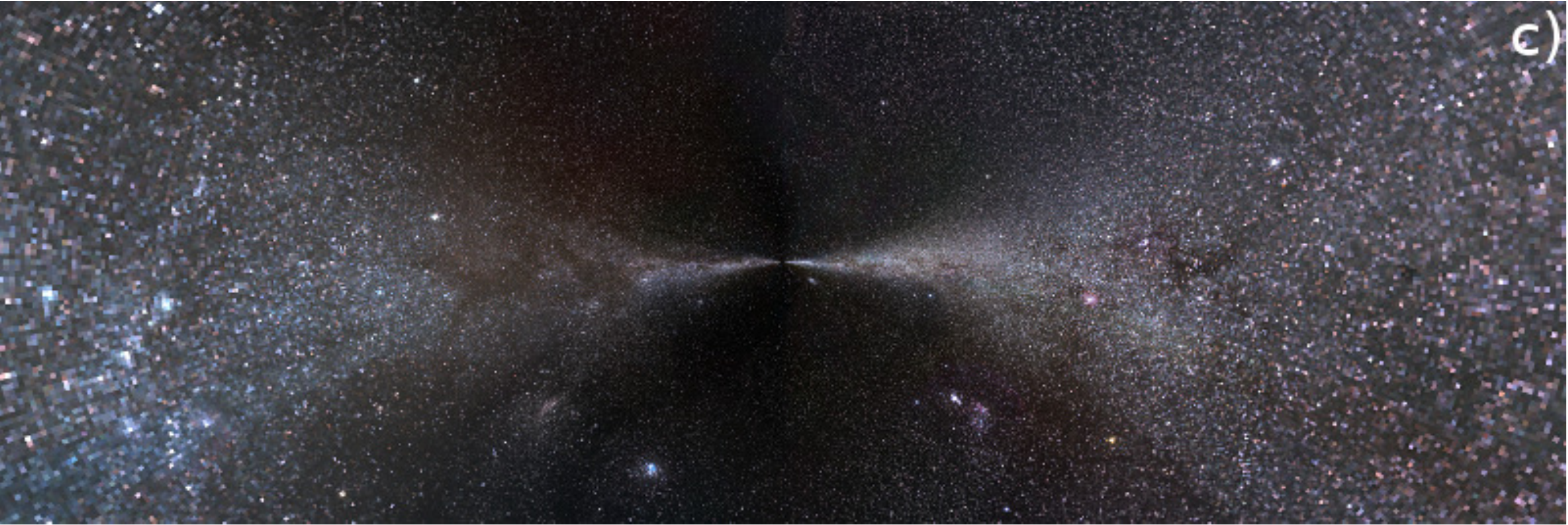}
 \caption{View of a static observer located at $x=10,y=0$ in the negative $x$-direction towards the approaching warp bubble that moves with velocity $v=2c$ and that crosses the origin $x=y=0$ at $t=0$. The panorama camera has $180\degree\times 60\degree$ field of view. The observation times are $t=\left\{0,3.5,3.98\right\}$ when the bubble is at $x_b=\left\{0,7,7.96\right\}$. The red ball is located at $x=0,y=3$. The Milky Way background is represented by a sphere with radius $r_{\mbox{max}}=200$.\cite{eso}}
 \label{fig:approach}
\end{figure}

\begin{figure}[ht]
 \centering
 \includegraphics[scale=0.55]{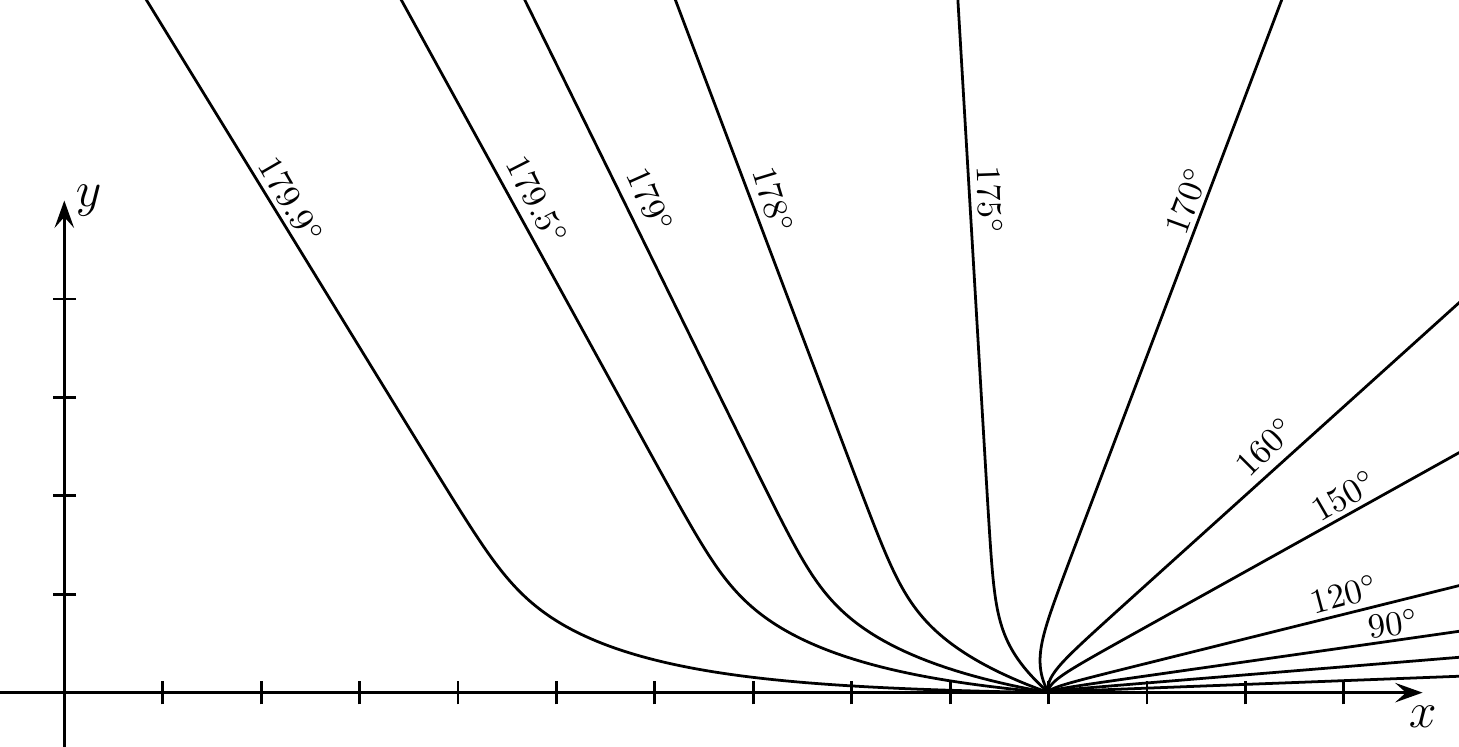}
 \caption{Incoming light rays for a static observer at $(t=3.98, x=10, y=0)$ and warp parameters $R=2$, $\sigma=1$, and $v=2c$.}
 \label{fig:towardsRays}
\end{figure}

\begin{figure}[ht]
 \centering
 \includegraphics[scale=0.72]{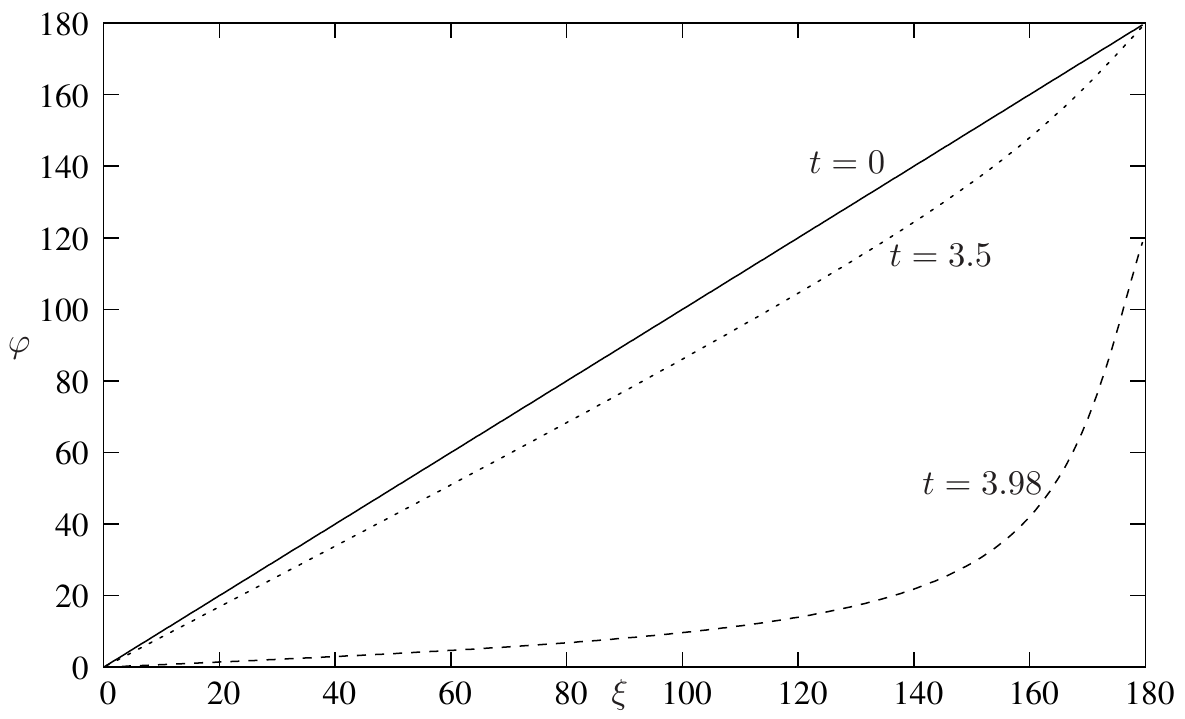}
 \caption{Angle $\varphi$ at $r_{\mbox{max}}=5\times 10^4$ with respect to the incident angle $\xi$ for a static observer at $x=10,y=0$ and $t=\left\{0,3.5,3.98\right\}$. The warp parameters read $R=2$, $\sigma=1$, and $v=2c$. The angle $\xi=180\degree$ corresponds to the direction of the approaching warp bubble.}
 \label{fig:twdangle}
\end{figure}

Consider a static observer located at $x=10,y=0$ and a warp bubble with $R=2$ and $\sigma=1$ that approaches the static observer with $v=2c$. At $t=0$, the bubble crosses the origin $x=0,y=0$. The observer, however, that looks towards the approaching warp bubble, will not recognize any significant image distortion, see figure \ref{fig:approach}(a). At $t=3.5$ (figure \ref{fig:approach}(b)), the warp bubble is located at $x_{\mbox{b}}=7$ and, thus, is already quite close to the observer that will see that some portion of the sky apparently shrinks in the direction of view. However, in the next few instances the view of the observer changes dramatically. Some parts of the sky in the direction of view disappear as a result of an apparent horizon like in the bridge observer example, see also Figs.~\ref{fig:towardsRays} and \ref{fig:twdangle}.

\begin{figure}[ht]
 \centering
 \includegraphics[scale=0.72]{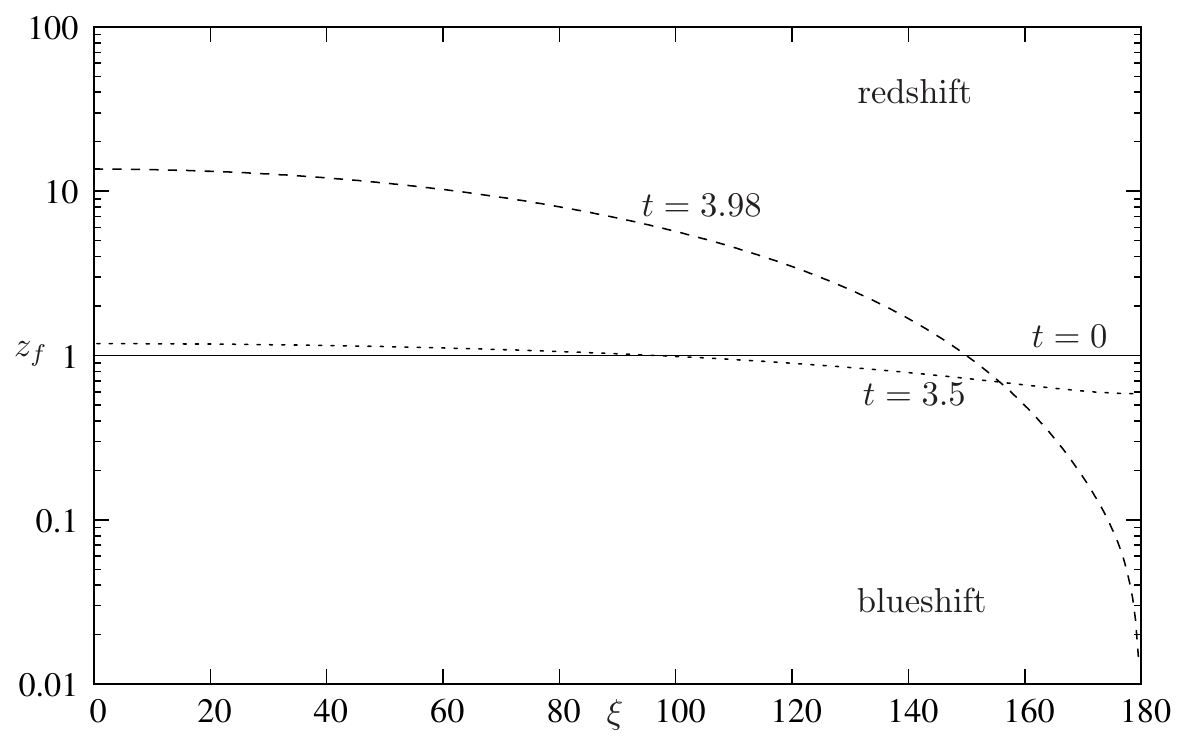}
 \caption{Frequency shift $1+z_f$ between $r_{\mbox{max}}=5\times 10^4$ and a static observer at $x=10,y=0$ and $t=\left\{0,3.5,3.98\right\}$. The warp parameters read $R=2$, $\sigma=1$, and $v=2c$.}
 \label{fig:twdz}
\end{figure}

\begin{figure}[ht]
 \centering
 \includegraphics[scale=0.72]{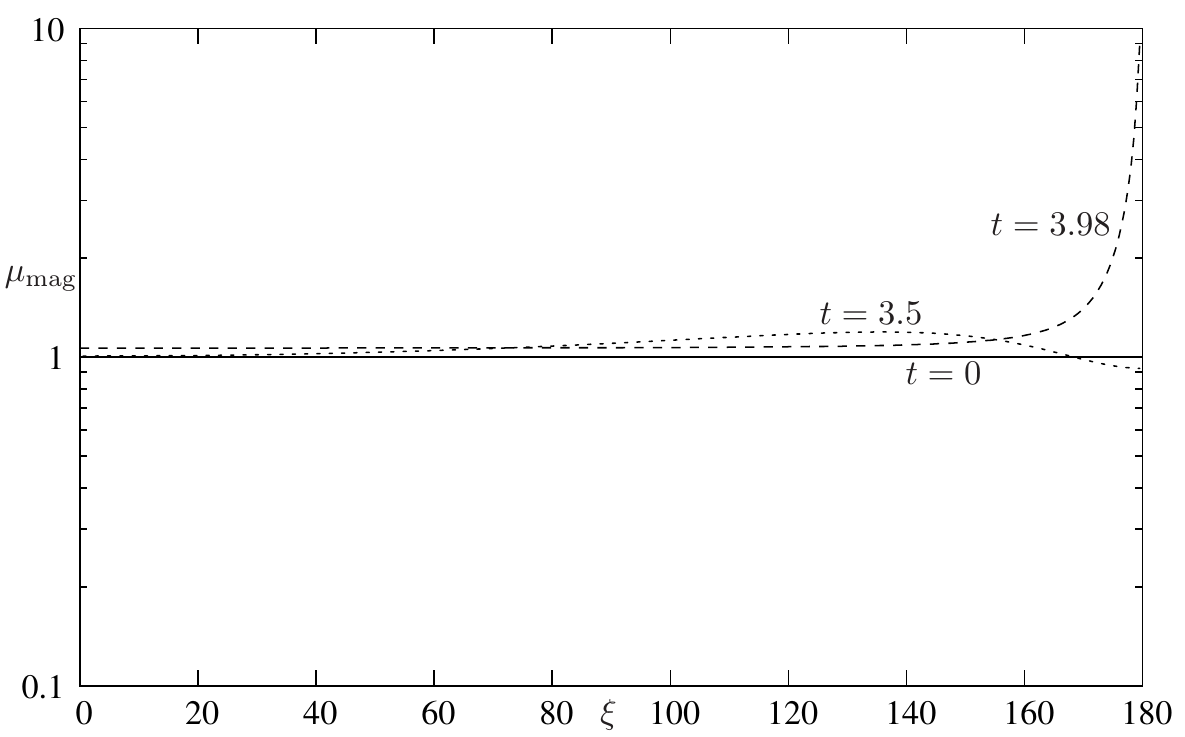}
 \caption{Lensing $\mu_{\mbox{mag}}$ between $r_{\mbox{max}}=5\times 10^4$ and a static observer at $x=10,y=0$ and $t=\left\{0,3.5,3.98\right\}$. The warp parameters read $R=2$, $\sigma=1$, and $v=2c$.}
 \label{fig:twdmu}
\end{figure}

The corresponding frequency shift and lensing diagrams are shown in Figs.~\ref{fig:twdz} and \ref{fig:twdmu}. Here, $\xi=180\degree$ represents the direction to the approaching warp bubble. For $t=0$, there is no significant frequency shift or magnification in any direction. For $t=3.98$ the warp bubble is located at $x_{\mbox{b}}=7.96$, however, the approaching warp bubble leads to a strong blueshift in the direction of the warp bubble and a redshift in the opposite direction. The lensing effect lets stars appear brighter especially in the direction of the approaching warp bubble.

%% ---------------------------------------------------------------------
\subsection{Warp bubble passing the observer}
\begin{figure}[ht]
 \centering
 \includegraphics[scale=0.52]{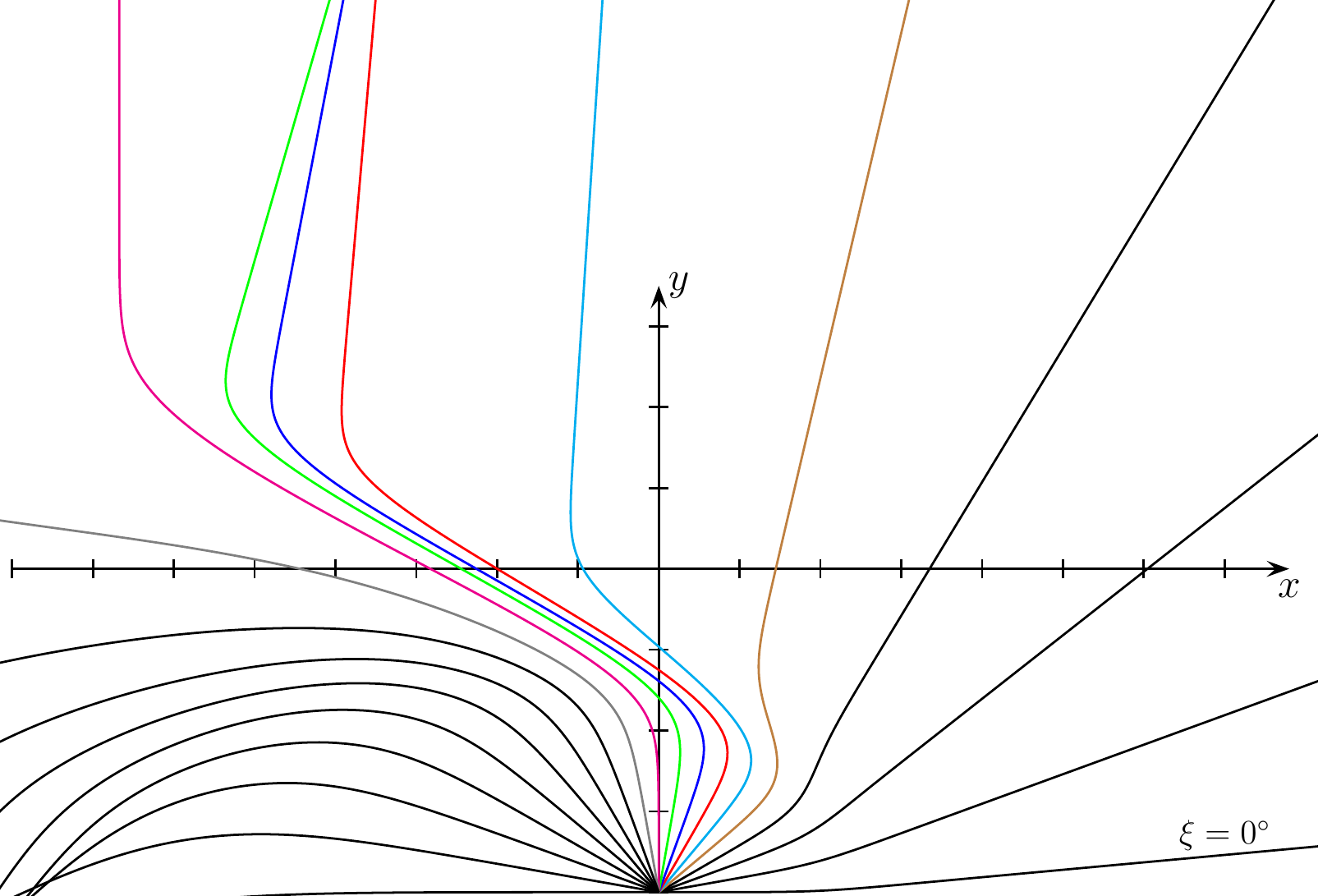}
 \caption{Incoming light rays for a static observer at $(t=3, x=0, y=-4)$ with incident angles $\xi=\left\{0\degree,10\degree,\ldots,180\degree\right\}$. The warp parameters read $R=2$, $\sigma=1$, and $v=2c$. The colored lines are only for better distinguishability.}
 \label{fig:incPass1}
\end{figure}

\begin{figure}[ht]
 \centering
 \includegraphics[scale=0.52]{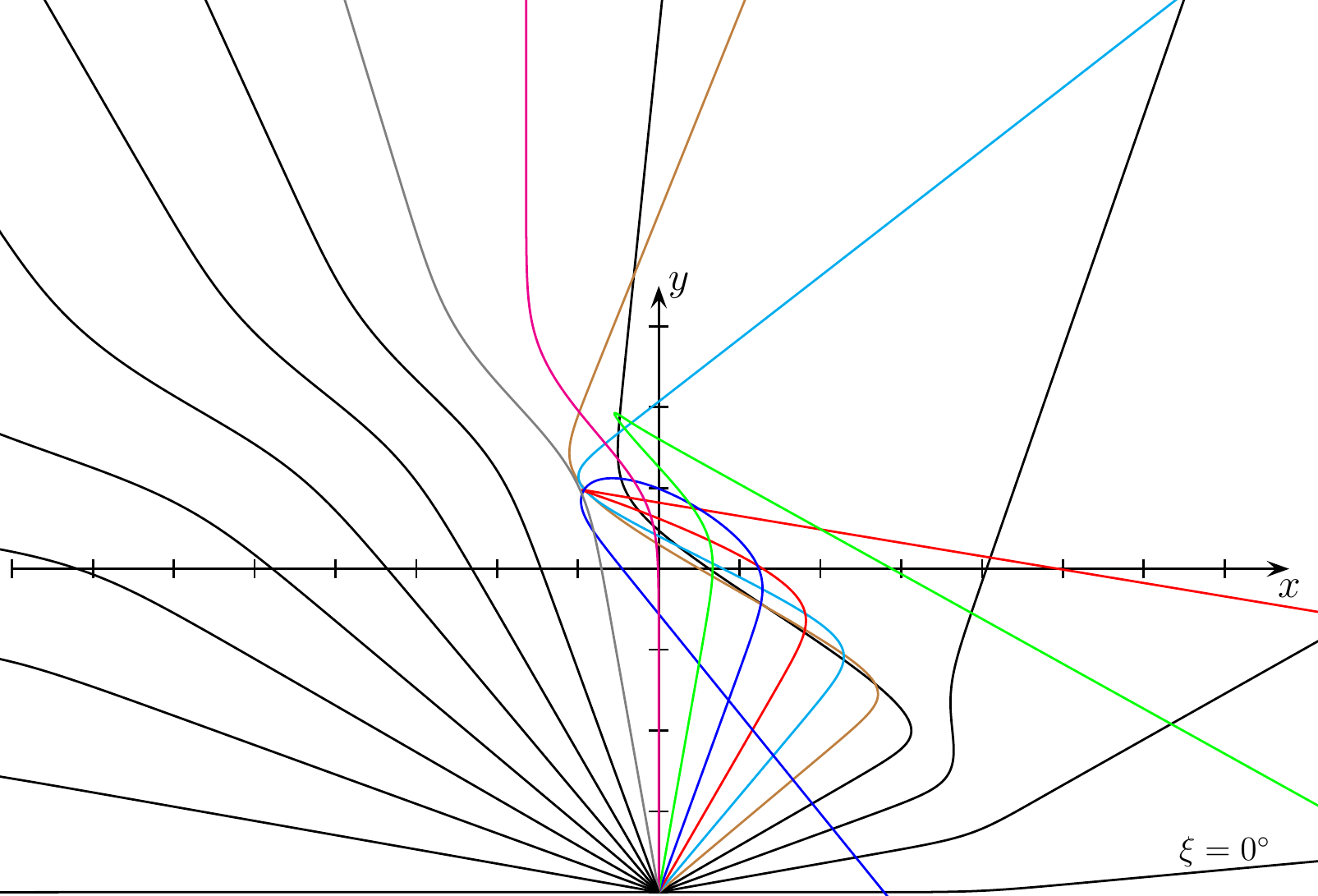}
 \caption{Incoming light rays for a static observer at $(t=6, x=0, y=-4)$ with incident angles $\xi=\left\{0\degree,10\degree,\ldots,180\degree\right\}$. The warp parameters read $R=2$, $\sigma=1$, and $v=2c$. The colored lines are only for better distinguishability.}
 \label{fig:incPass2}
\end{figure}

\begin{figure}[ht]
 \centering
 \includegraphics[scale=0.3]{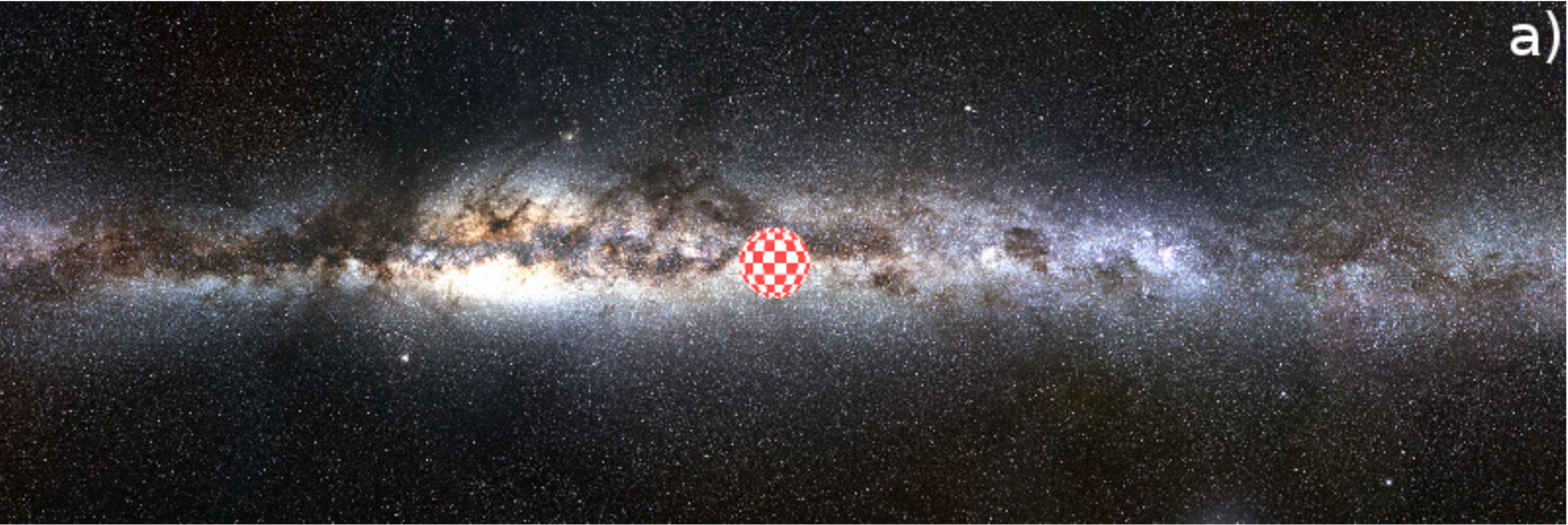}\\
 \includegraphics[scale=0.3]{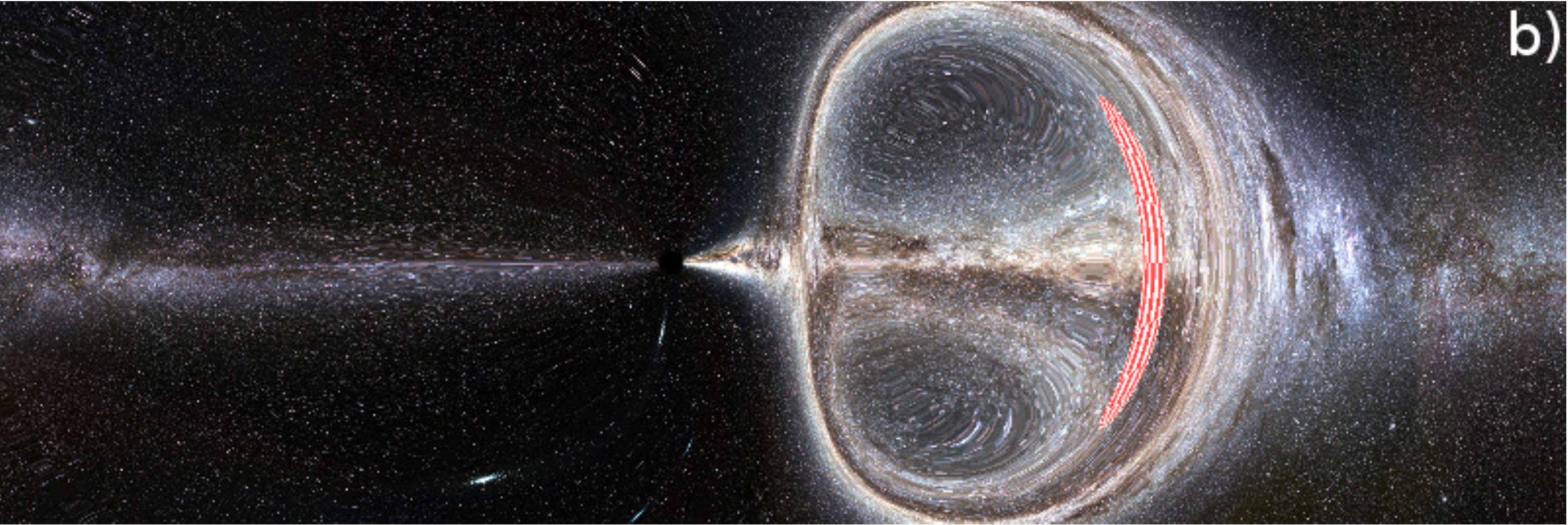}\\
 \includegraphics[scale=0.3]{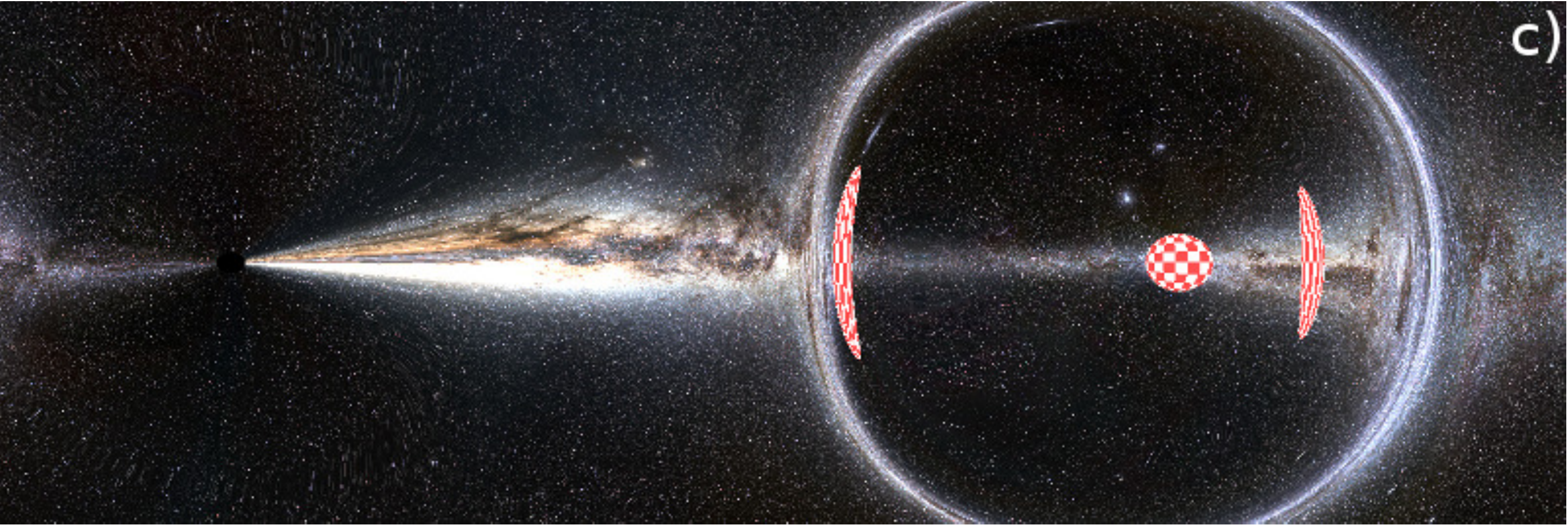}\\
 \includegraphics[scale=0.3]{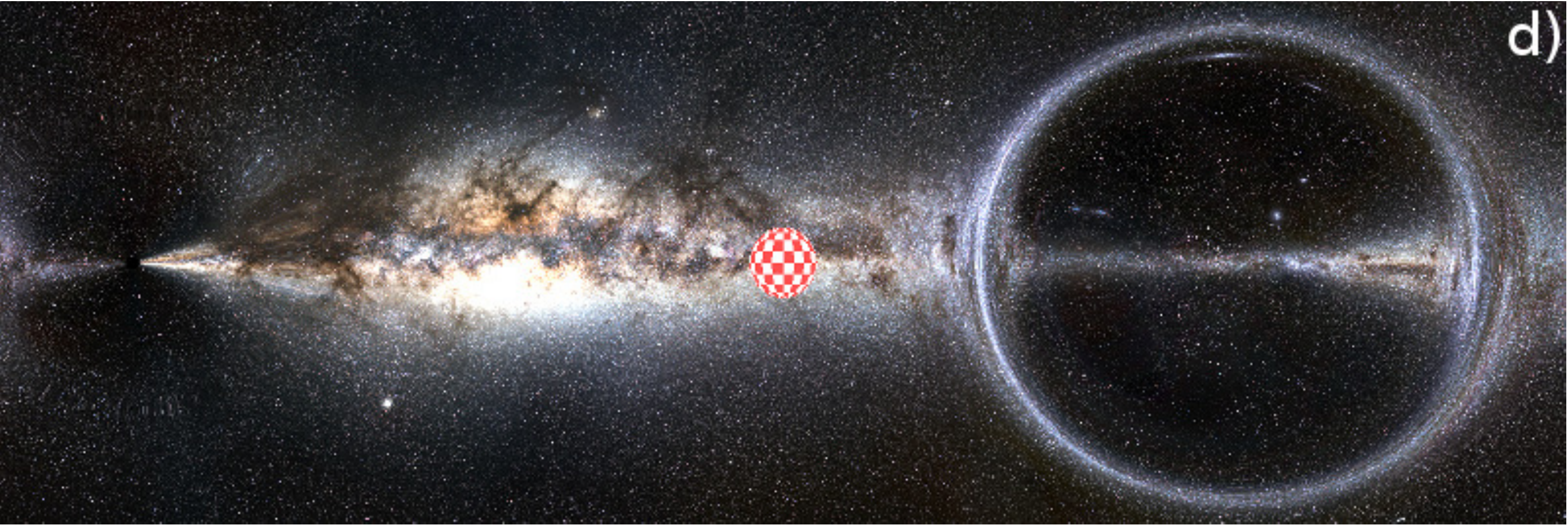}
 \caption{View of a static observer located at $x=0,y=-4$ in the positive $y$-direction for $t=0$ (a), $t=3$ (b), $t=6$ (c), and $t=9$ (d). The warp parameters read $R=2$, $\sigma=1$, and $v=2c$. The panorama camera has $180\degree\times 60\degree$ field of view. The ball is located at $x=0,y=3$. The Milky Way background is represented by a sphere with radius $r_{\mbox{bg}}=200$.}
 \label{fig:passImages}
\end{figure}

Consider a static observer located at $x=0,y=-4$ and a warp bubble with parameters $R=2$, $\sigma=1$ that passes the observer with $v=2c$. At $t=0$, the bubble crosses the origin $x=0,y=0$. Figures~\ref{fig:incPass1} and \ref{fig:incPass2} show several representing light rays that reach the static observer at times $t=3$ and $t=6$ with incident angles $\xi=\left\{0\degree,10\degree,\ldots,180\degree\right\}$. The corresponding views for a panorama camera with $180\degree\times 60\degree$ field of view heading towards the origin are shown in figure \ref{fig:passImages}. Here, the Milky Way background is represented by a sphere with radius $r_{\mbox{max}}=200$, and the checkered ball of radius $r_{\mbox{ball}}=0.5$ is located at $x=0,y=3$.

\begin{figure}[ht]
 \centering
 \includegraphics[scale=0.72]{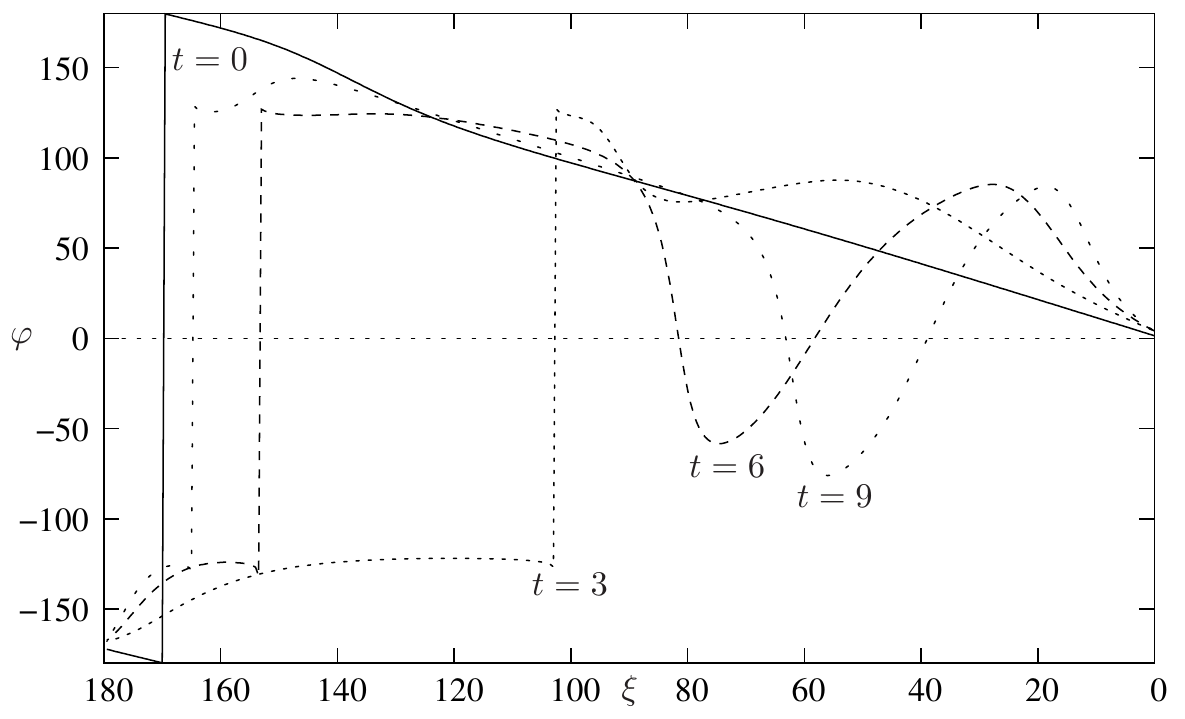}
 \caption{Angle $\varphi$ at $r_{\mbox{max}}=200$ with respect to incident angle $\xi$ for $t=\left\{0,3,6,9\right\}$ and an observer located at $x=0,y=-4$. The warp parameters read $R=2$, $\sigma=1$, and $v=2c$.}
 \label{fig:passAngle}
\end{figure}

The three phantom images of the ball (figure \ref{fig:passImages}(c)) follow from light rays with incident angles $\xi\approx\left\{28.8\degree,45\degree,83.2\degree\right\}$. The distortion of the Milky Way background can be understood by means of the relation between the intersection angles $\varphi$ of the light ray with the background sphere and the incident angles $\xi$, see figure \ref{fig:passAngle}. The discontinuities reflect the apparent horizons at $\xi_{\mbox{hor}}\approx\left\{171.3\degree,102.8\degree,153.6\degree,164.8\degree\right\}$ for the corresponding observation times $t=\left\{0,3,6,9\right\}$. Since the mapping $\varphi\mapsto\xi$ is non-injective, some parts of the sky appears more than once. For example, the point $(r=r_{\mbox{max}}, \varphi=50\degree)$ appears three times at observation time $t=9$ under the incident angles $\xi\approx\left\{10.82\degree,30.83\degree,67.85\degree\right\}$. 

What can also be read from figure \ref{fig:passAngle} is that, after the warp bubble has passed the observer, there is another distortion region that apparently moves in the negative $x$-direction. This secondary distortion region can be easily understood. Due to the finite speed of light, light rays that have already traversed the warp bubble region at earlier times now reach the observer. Thus, by increasing time, the observer receives ever earlier light rays.

\begin{figure}[ht]
 \centering
 \includegraphics[scale=0.72]{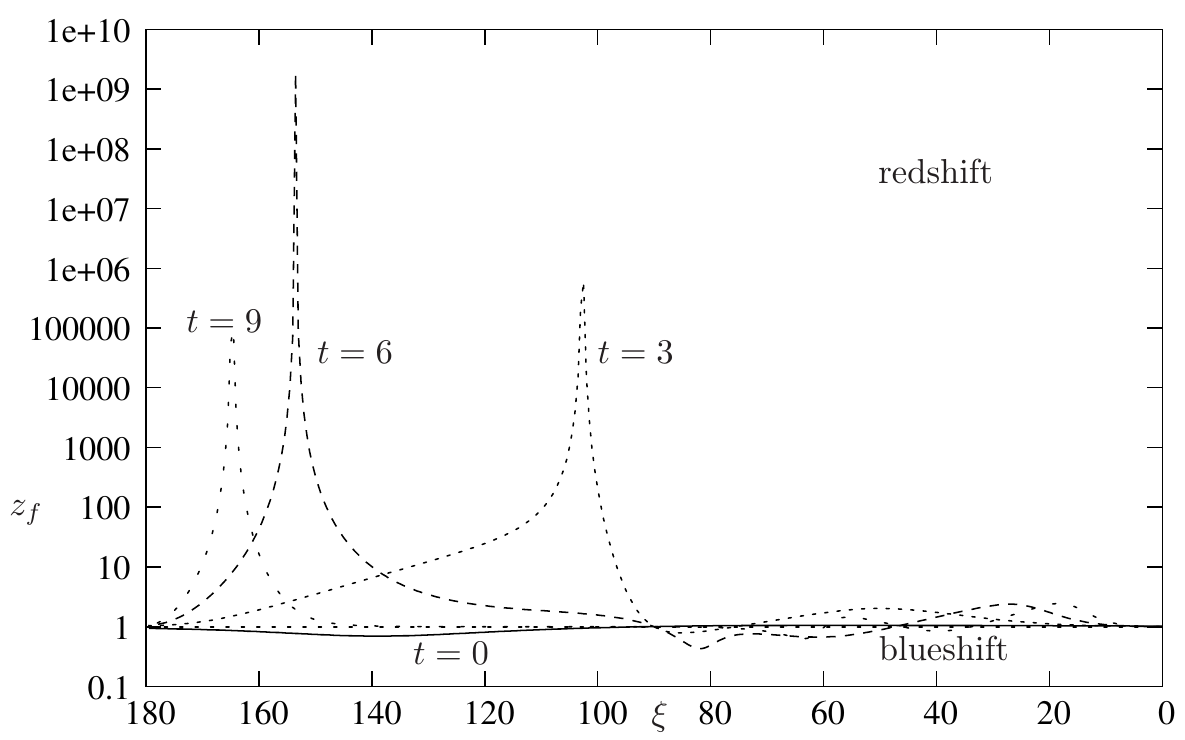}
 \caption{Frequency shift $1+z_f$ between $r_{\mbox{max}}=200$ and the observer located at $x=0,y=-4$ with respect to incident angle $\xi$ for $t=\left\{0,3,6,9\right\}$. The warp parameters read $R=2$, $\sigma=1$, and $v=2c$.}
 \label{fig:passZ}
\end{figure}

Figure~\ref{fig:passZ} shows the frequency shift for the observation times $t=\left\{0,3,6,9\right\}$. At $t=0$, the frequency shift is negligible. However, when the warp bubble has passed the origin, there is some blue- and redshift from the primary distortion region $(\xi>90\degree)$, whereas in the secondary region there is a strong redshift close to the apparent horizon directions.

\begin{figure}[ht]
 \centering
 \includegraphics[scale=0.72]{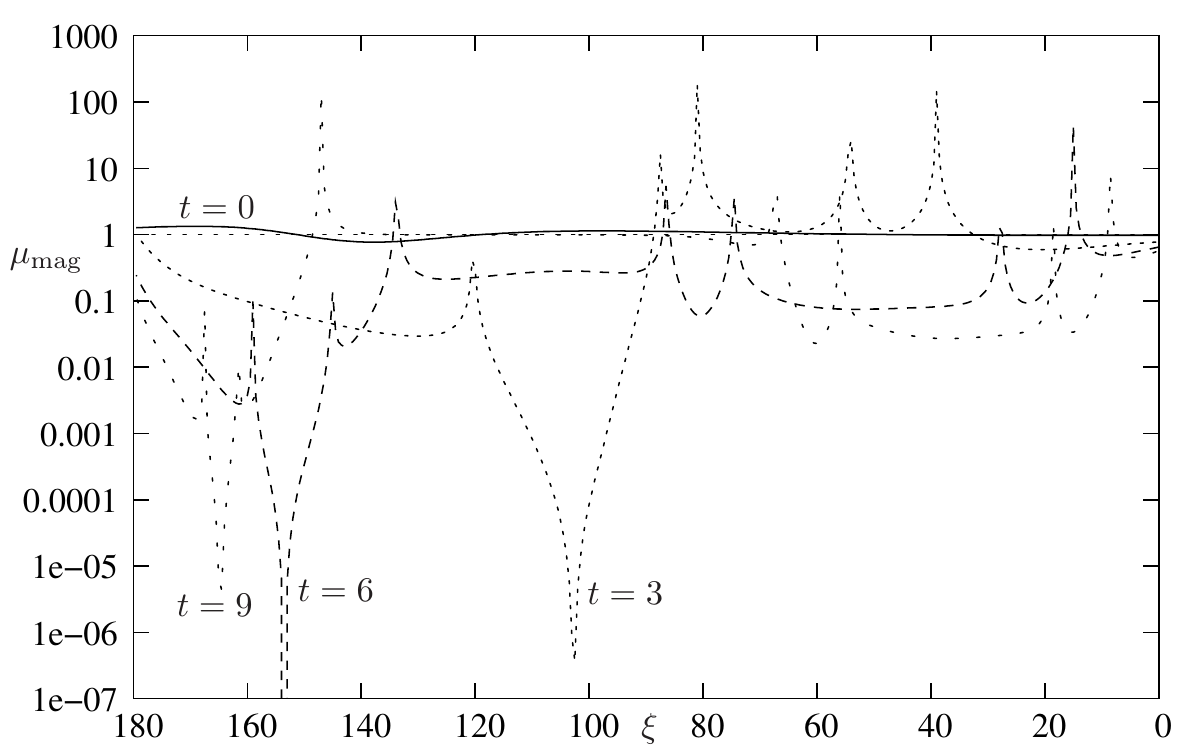}
 \caption{Magnification $\mu_{\mbox{mag}}$ between $r_{\mbox{max}}=200$ and the observer located at $x=0,y=-4$ with respect to incident angle $\xi$ for $t=\left\{0,3,6,9\right\}$. The warp parameters read $R=2$, $\sigma=1$, and $v=2c$.}
 \label{fig:passMu}
\end{figure}

Figure~\ref{fig:passMu} shows the lensing effect for the observation times $t=\left\{0,3,6,9\right\}$. Similar to the frequency shift, the magnification at $t=0$ is negligible. However, when the warp bubble passes the observer, there are strong magnifications at both sides of each bubble rim. In between, stars would appear dimmed. In the regions where the redshift is dominant, the lensing effects fades out the star light.
% -----------------------------------------------------------------
%                      interactive flight
% -----------------------------------------------------------------
\section{Interactive flight}\label{sec:vis}
In contrast to the usual graphical representation of the visual appearance of a star field in science fiction movies with its elongated stars, the actual view of such a star field as seen from within a warp bubble is completely different. To demonstrate what could be really seen, we have developed a Java application with which the user can travel through the stars of the Hipparcos catalogue either with relativistic speeds in the Minkowskian spacetime or with warp speed in the Alcubierre metric.

In the following subsection, we briefly describe the relativistic star flight simulator \texttt{JRelStarFlight} and its graphical user interface. Technical details about the implementation and the calculation of the physical quantities are discussed in subsection \ref{subsec:techdetails}.

%% -----------------------------------------------------------
\subsection{JRelStarFlight}\label{subsec:RSFgui}
\begin{figure}[ht]
 \centering
 \includegraphics[scale=0.3]{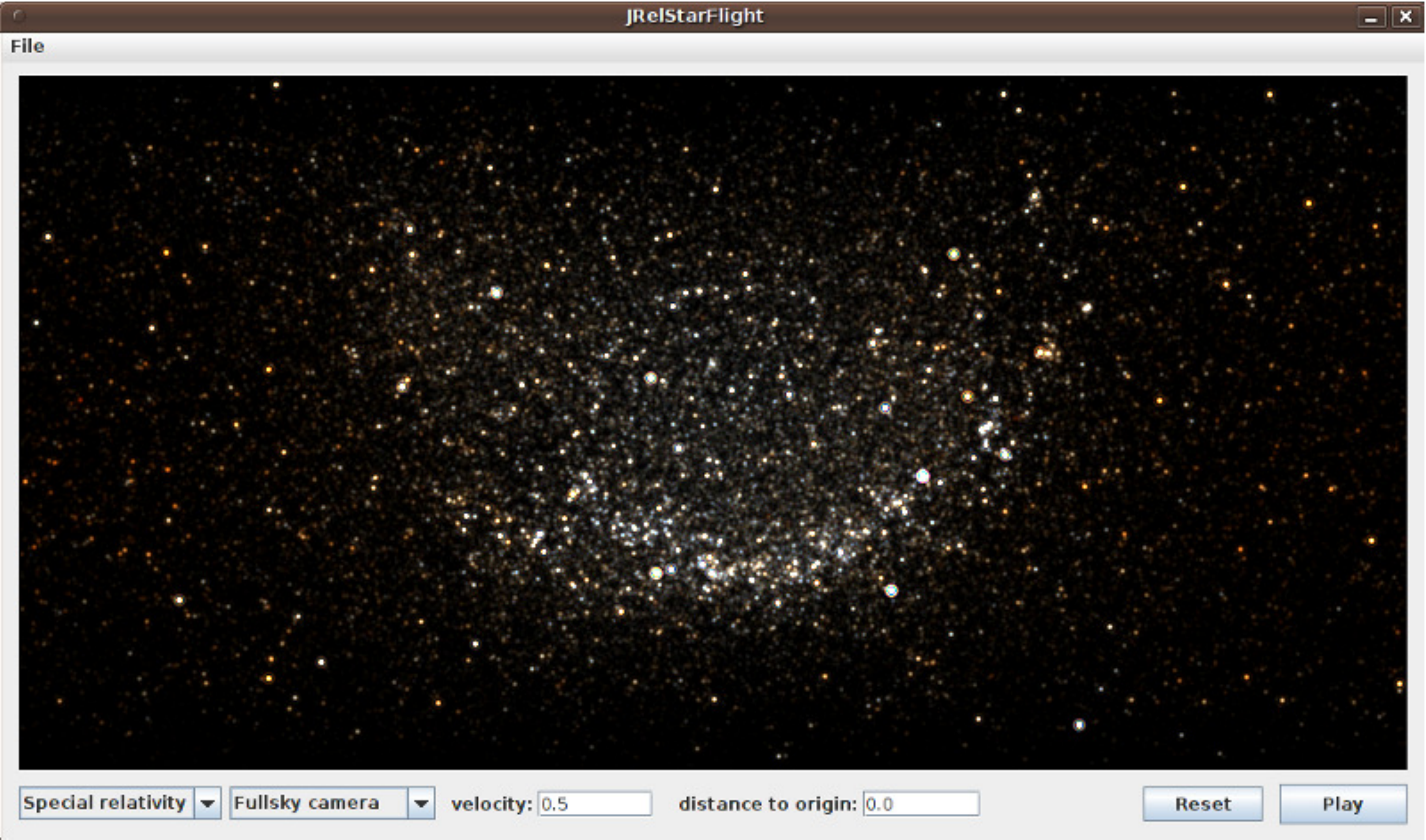}
 \caption{Screenshot of the Java application {\it JRelStarFlight}.}
 \label{fig:screenshot}
\end{figure}

The graphical user interface of our relativistic flight simulator \texttt{JRelStarFlight} is shown in figure \ref{fig:screenshot}. When the application is launched, the observer is at rest in a Minkowskian spacetime and uses a full-sky panorama camera. To toggle between special relativistic and warp flight, use the drop-down menu in the lower left corner. There are two camera modes: the full-sky camera maps the whole sky from spherical coordinates onto a rectilinear like grid; the pinhole camera acts as a normal camera with $50\degree$ vertical field of view.

Moving the mouse in the OpenGL window with the left mouse button pressed, the viewing direction and, thus, the direction of motion can be changed. The current velocity $\beta=v/c$ can be modified by means of the input field `beta'. Toggling the `play' button lets the observer move or stop. The current position can also be set with the input field `distance to origin'. Please note that when the observer has moved away from the origin, the mouse rotation acts as if the observer has moved in the newly selected direction ab initio.

%% -----------------------------------------------------------
\subsection{Technical details}\label{subsec:techdetails}
The apparent position of a point-like star as seen within the warp bubble depends on how a light ray of this star is influenced by the curved spacetime. For the visual appearance, we also have to take the frequency shift and the lensing effect into account. 

For the actual visualization of the stars we adopt the method by M{\"u}ller and Weiskopf~\cite{mueller2010b}. As database we use the Hipparcos star catalogue~\cite{hipparcos} and assign a Planck spectrum to each star with temperature $T_{\mbox{star}}$ calculated by Reed's empirical law~\cite{reed}. The frequency-shifted spectrum is again a Planck spectrum but at the different temperature $T_{\mbox{obs}}=T_{\mbox{star}}/(1+z_f)$. The resulting apparent visual magnitude of a star is determined by the magnification factor $\mu_{\mbox{mag}}$ and the integral over the Planck spectrum in the visual wavelength domain. A finite eye or telescope aperture yields a Fraunhofer diffraction pattern.

Since the star distances are in terms of light-years, we use years as unit of time. With $25$ frames per second and a step size of $\Delta t=0.4y$ per frame, one second of visualization time corresponds to $10$ years of simulation time.

%% -----------------------------------------------------------
\subsection{Implementation details}\label{subsec:impldetails}
Before running the Java application we have to generate the two-dimensional lookup table that stores the observation angle $\xi$, the frequency shift $z_f$, and the magnification factor $\mu_{\mbox{mag}}$ for each asymptotic light direction $\varphi$ and velocity $v$ of the warp bubble. For that, we integrate the geodesic equation, the parallel transport of the Sachs basis vectors, and the Jacobian equations from the observer back in time until the geodesic reaches the sphere with radius $r_{\mbox{max}}=5\times 10^4$. Hence, we obtain a relation between observation angles $\xi$ and intersection angles $\varphi$ with the sphere, which must be inverted at the end of the calculations.

For our simulator, we use the warp parameters $R=2,\sigma=1$ and a lookup table with resolution $4096\times 248$. While the angle $\varphi$ is sampled linearly from $0$ to $\pi$, the velocity in the row number `$\mbox{row}$' is given by
\begin{equation}
  v = 10^{(\mbox{row}+1)/248}-1.
\end{equation}
Thus, the minimum velocity $v_{\mbox{min}}\approx 0.0093$ is stored in $\mbox{row}=0$ and the maximum velocity $v_{\mbox{max}}=9$ in $\mbox{row}=247$.

The rendering of the stars is realized using a Java implementation of the open graphics library OpenGL and the shading language GLSL~\cite{opengl}. In the vertex shader, we use the above explained lookup table to calculate the apparent positions and frequency-shifted Planck temperatures of the stars. In the fragment shader, the Planck temperatures are mapped to the precalculated color and Fraunhofer diffraction pattern.

% -----------------------------------------------------------------
%                      timelike geodesics
% -----------------------------------------------------------------
\section{Timelike geodesics}\label{sec:timelikeGeodesics}
So far, we considered only the influence of the warp metric on light rays. Equally interesting, however, is the influence on massive particles as we will discuss in this section. As with the null geodesics, we use the non-affinely parametrized geodesic equation with the coordinate time as parameter. 

%% ------------------------------------------------------
\subsection{Particles from the bridge}
\begin{figure}[ht]
 \centering
 \includegraphics[scale=0.5]{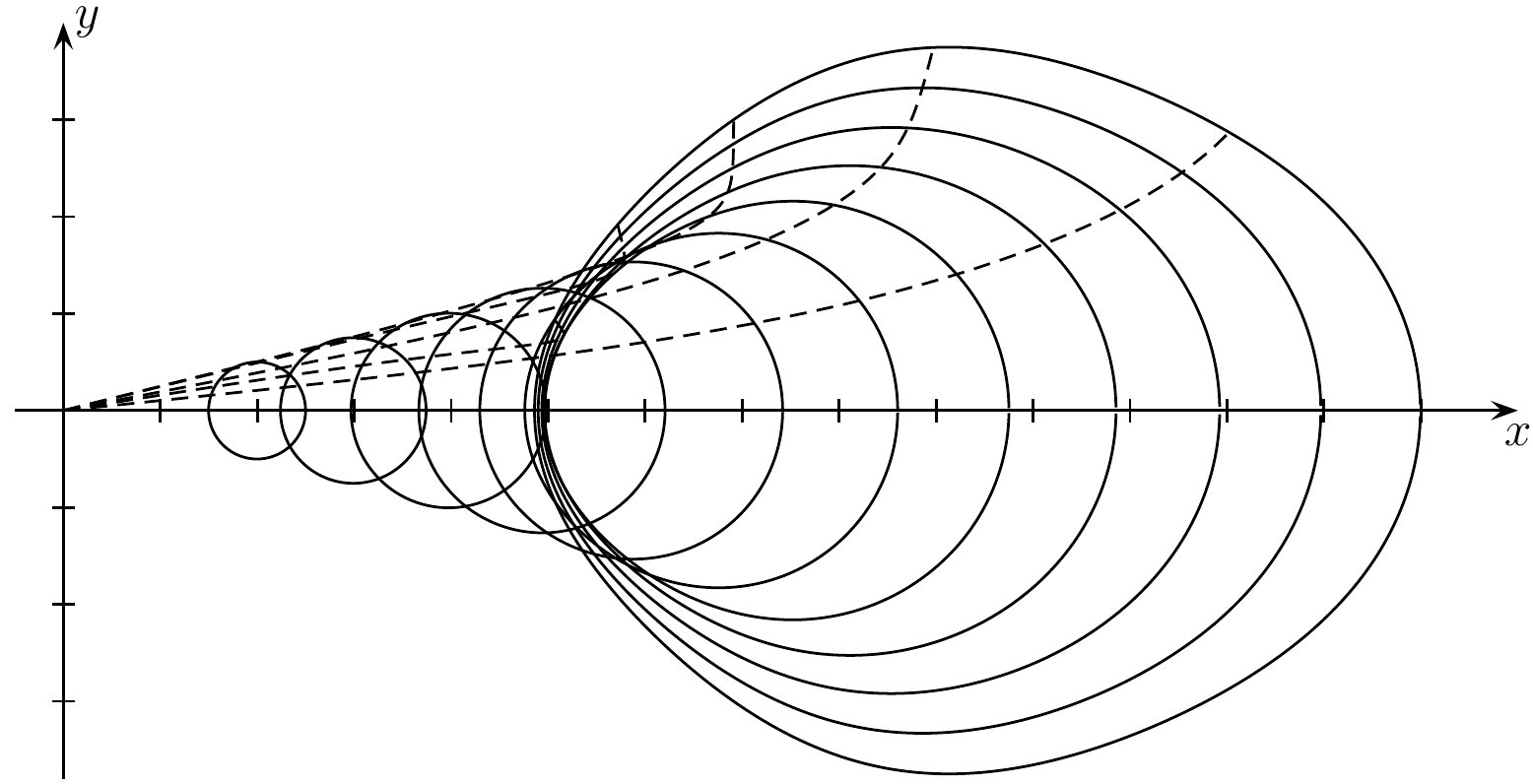}
 \caption{Particles emitted radially from the bridge with initial local velocity $v_{\mbox{part}}=0.5c$ after $\Delta t=\left\{1.0,1.5,\ldots,6.0\right\}$ (solid curves). The dashed lines correspond to particle trajectories for initial directions $\xi=\left\{0\degree,30\degree,\ldots,150\degree\right\}$. The warp parameters read $R=2$, $\sigma=1$, $v=2c$.}
 \label{fig:partBridge}
\end{figure}

A particle at the center of the warp bubble with zero initial velocity will stay there for ever irrespective of the worldline $x(t)$ of the bubble. So, let us consider particles with initial local velocity $v_{\mbox{part}}$ with respect to the comoving reference frame and the corresponding four-velocity 
\begin{equation}
 \mathbf{u}=c\gamma\left[\mathbf{e}_{(0)}+\beta\left(\cos\xi\mathbf{e}_{(1)}+\sin\xi\mathbf{e}_{(2)}\right)\right] = u^{\mu}\partial_{\mu},
\end{equation}
where $\beta=v_{\mbox{part}}/c$ and $\gamma=1/\sqrt{1-\beta^2}$. Figure~\ref{fig:partBridge} shows particles that were emitted radially from the bridge with initial local velocity $v_{\mbox{part}}=0.5c$. 
Each solid line represents all particles that have travelled a specific coordinate time $\Delta t$. The dashed lines represent particle trajectories for a few initial directions $\xi$.

In the first few seconds, the particles still move within the warp bubble and are carried along with it. Depending on their initial direction $\xi$, they can leave the warp bubble after some time $\Delta t$.

\begin{figure}[ht]
 \centering
  \includegraphics[scale=0.72]{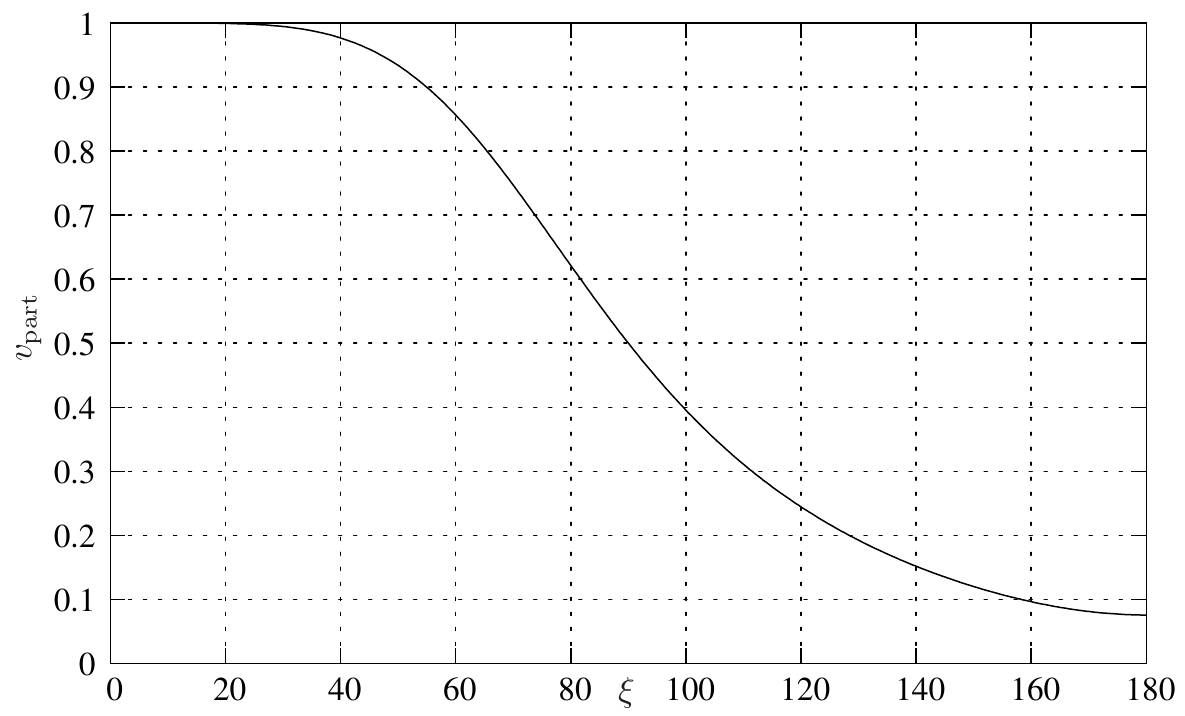}
  \caption{Particle velocities $v_{\mbox{part}}(t=1000)/c$ far outside the bubble's sphere of influence after $\Delta t$ that were emitted radially from the bridge with initial local velocity $v_{\mbox{part}}(\tau=0)=0.5c$. Particles with initial angles $\xi\lesssim 5\degree$ are still in the sphere of influence of the warp bubble.}
  \label{fig:partBridgeVel}
\end{figure}

To measure the current velocity of a particle during this period, we can use only a valid observer, which is an observer represented by a comoving local tetrad. Such an observer will always measure a velocity less than the speed of light. Figure~\ref{fig:partBridgeVel} shows the velocities of the particles after $\Delta t=1000$ when all of them are far away from the sphere of influence of the warp bubble. Similar to null geodesics, a particle with initial direction $\xi=90\degree$ is only displaced by the warp bubble and the velocity keeps unchanged. All other particles have either a higher or a lower velocity depending on the initial angle $\xi$.

%% ------------------------------------------------------
\subsection{Particles injected from outside}
\begin{figure}[ht]
 \centering
 \includegraphics[scale=0.63]{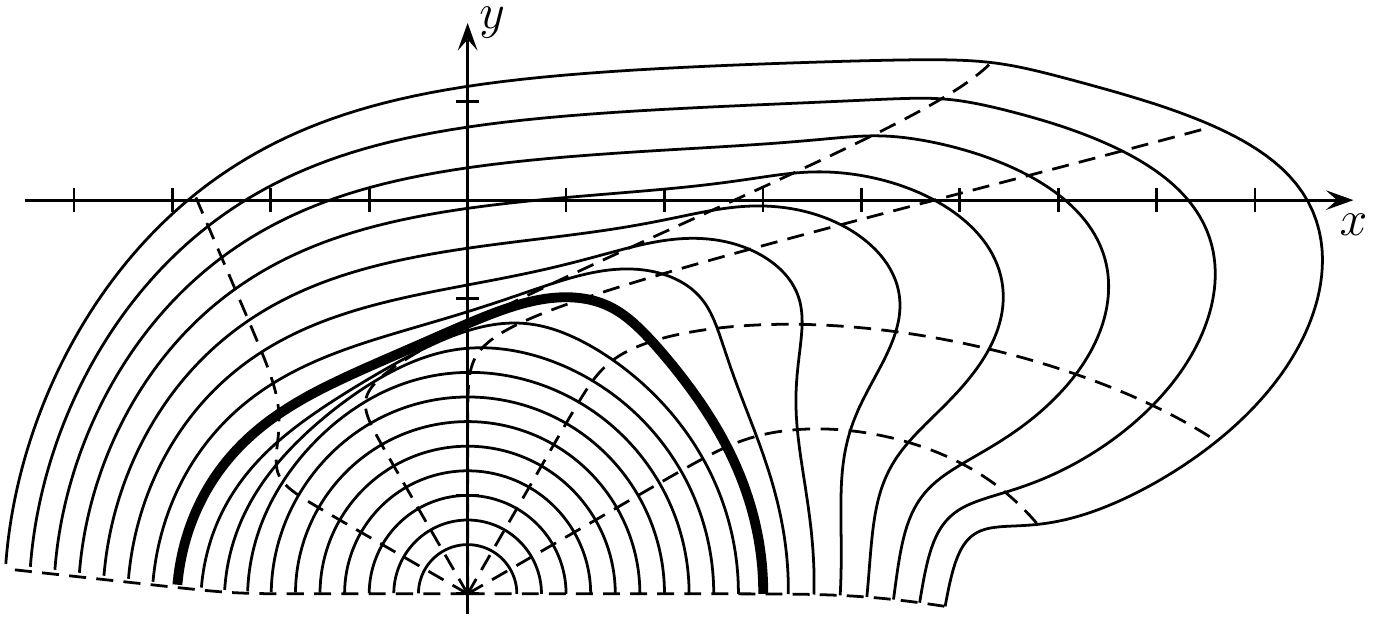}
 \caption{The solid lines represent particles that start with $\beta=0.5$ from the position $(t=-6, x=0, y=-4)$. The lines are separated by $\Delta t=0.5$. The thick solid line indicates $t=0$. The warp parameters read $R=2$, $\sigma=1$, $v=2c$. The dashed lines correspond to timelike geodesics with initial angles $\xi=0\degree,30\degree,\ldots,180\degree$.}
 \label{fig:particleFront}
\end{figure}

\begin{figure}[ht]
 \centering
 \includegraphics[scale=0.63]{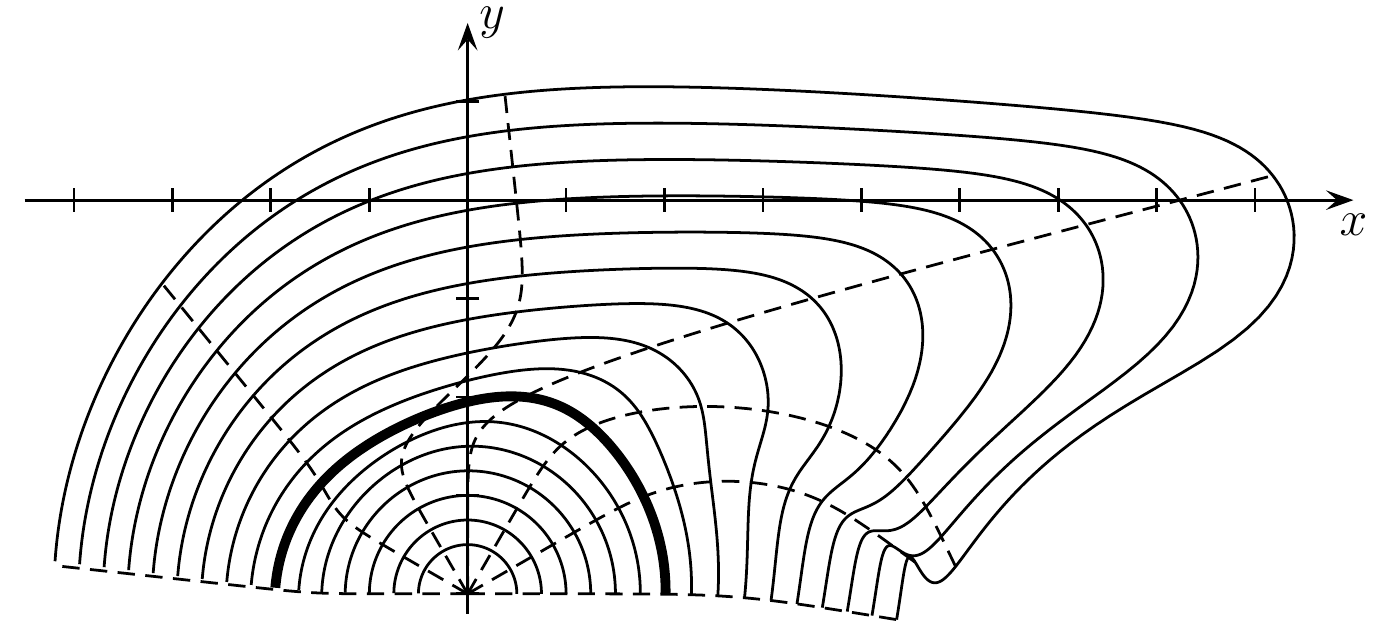}
 \caption{The same situation as in figure \ref{fig:particleFront} but with initial position $(t=-4, x=0, y=-4)$.}
 \label{fig:particleFrontTwo}
\end{figure}

Consider a static observer located at $x=0,y=-4$ that emits particles with local velocity $v_{\mbox{part}}=0.5c$ in the directions $\xi\in[0\degree,180\degree]$ at coordinate time $t$. The corresponding four-velocity $\mathbf{u}$ with respect to the static reference frame reads
\begin{equation}
 \mathbf{u}=c\gamma\left[\mathbf{\hat{e}}_{(0)}+\beta\left(\cos\xi\mathbf{\hat{e}}_{(1)}+\sin\xi\mathbf{\hat{e}}_{(2)}\right)\right] = u^{\mu}\partial_{\mu},
\end{equation}
where $\beta=v_{\mbox{part}}/c$ and $\gamma=1/\sqrt{1-\beta^2}$. The solid lines of figures~\ref{fig:particleFront} and \ref{fig:particleFrontTwo} show particles that were emitted at coordinate times $t=-6$ or $t=-4$, respectively, and that have travelled some time $\Delta t$. The thick solid lines correspond to $t=0$. The dashed lines represent particle trajectories for a few initial directions $\xi$.

Depending on the emission time, the particles reach the sphere of influence of the warp bubble under different impact angles. Hence, they will be deflected, carried along, and accelerated or decelerated in different ways.

%% ------------------------------------------------------
\subsection{Particle field initially at rest}
Consider a field of $5\times 6=30$ particles represented by tiny balls that are initially at rest with axes in the positive $x$ direction. The initial position of particle $(i,j)$ is given by $x_i=-3+i$, $y_j=-4+j$.  At the beginning of the simulation, $t=-3$, the warp bubble with parameters $R=2,\sigma=1, v=2c$ is located at $x=-6,y=0$. To determine the influence of the warp bubble on this particle field, we calculate the parallel transport of each particle. For that, we have to integrate the geodesic equation with initial conditions $x^{\mu}\big|_{\tau=0}=\left(-3,x_i,y_i,0\right)$, $dx^{\mu}/d\tau\big|_{\tau=0}=c\mathbf{\hat{e}}_{(0)}$ together with the parallel-transport equation
\begin{equation}
 \frac{dp^{\mu}}{d\tau}+\Gamma_{\nu\rho}^{\mu}\frac{dx^{\nu}}{d\tau}\frac{dp^{\rho}}{d\tau}=0,
\end{equation}
where the four-vector $p^{\mu}\big|_{\tau=0}=(0,1,0,0)$ describes the orientation of the particle. Like the geodesic equation, we first convert the parallel-transport equation into its non-affinely parametrized form (compare the parallel transport of the Sachs basis vectors explained in App.~\ref{app:sachsjac}).

\begin{figure}[ht]
 \centering
 \includegraphics[scale=0.45]{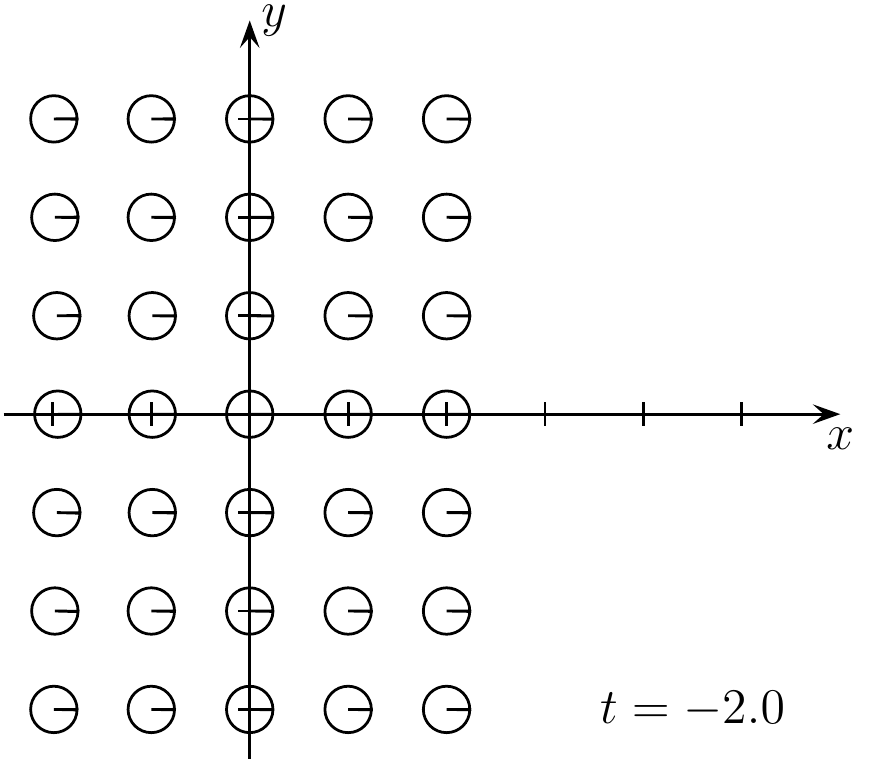}
 \includegraphics[scale=0.45]{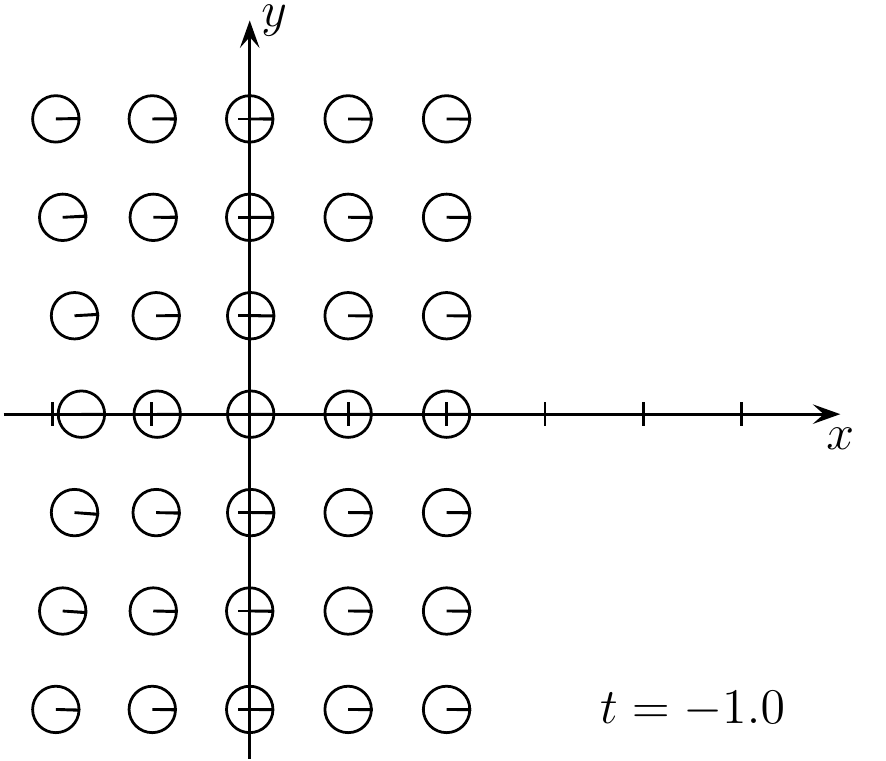}\\[0.3em]
 \includegraphics[scale=0.45]{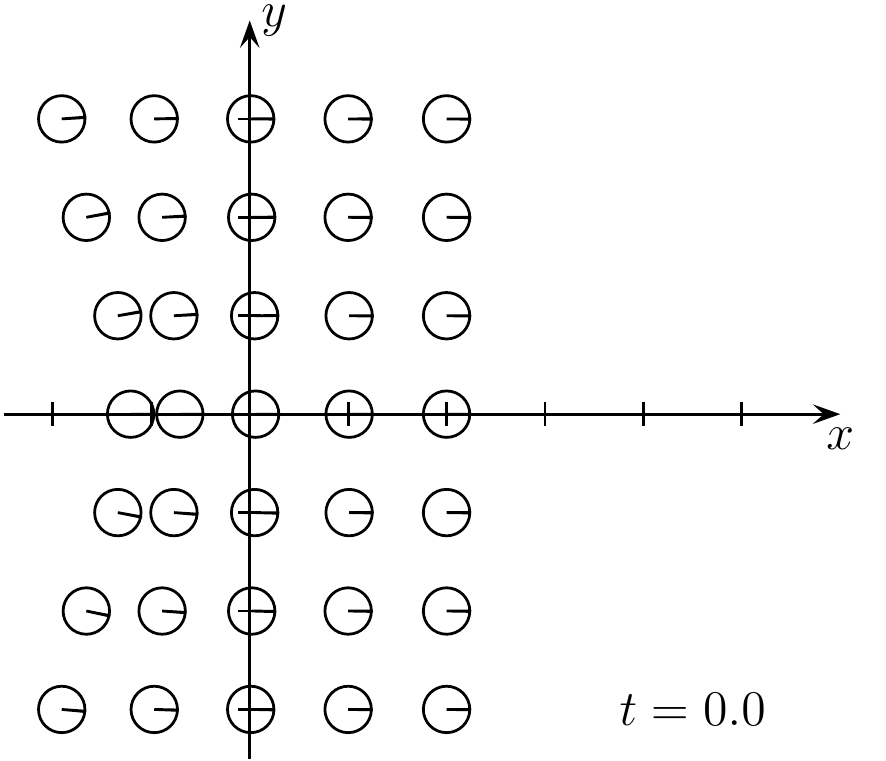}
 \includegraphics[scale=0.45]{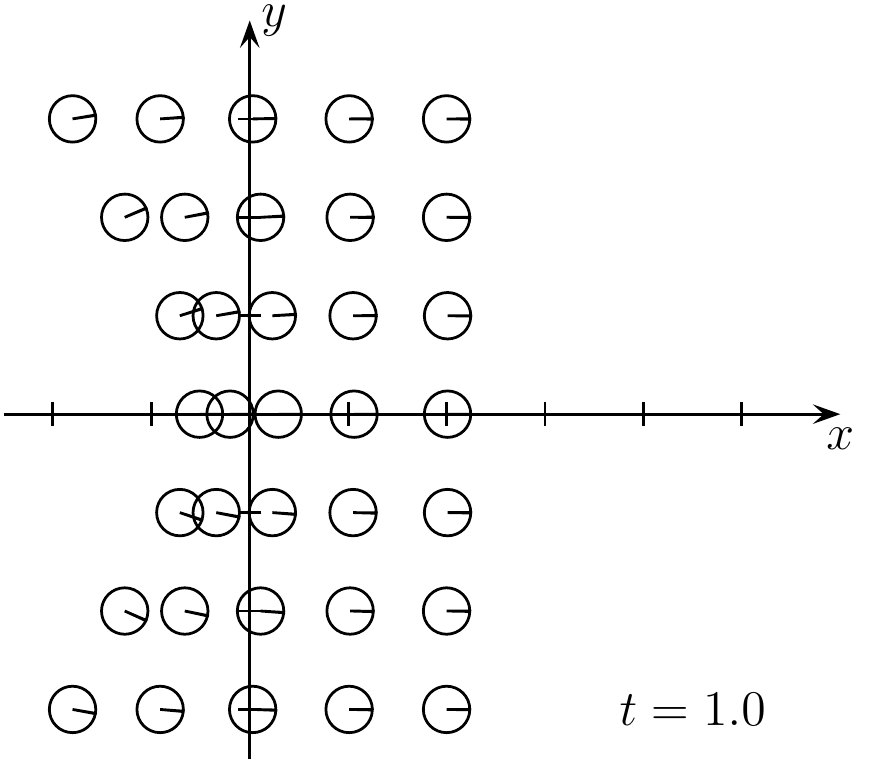}\\[0.3em]
 \includegraphics[scale=0.45]{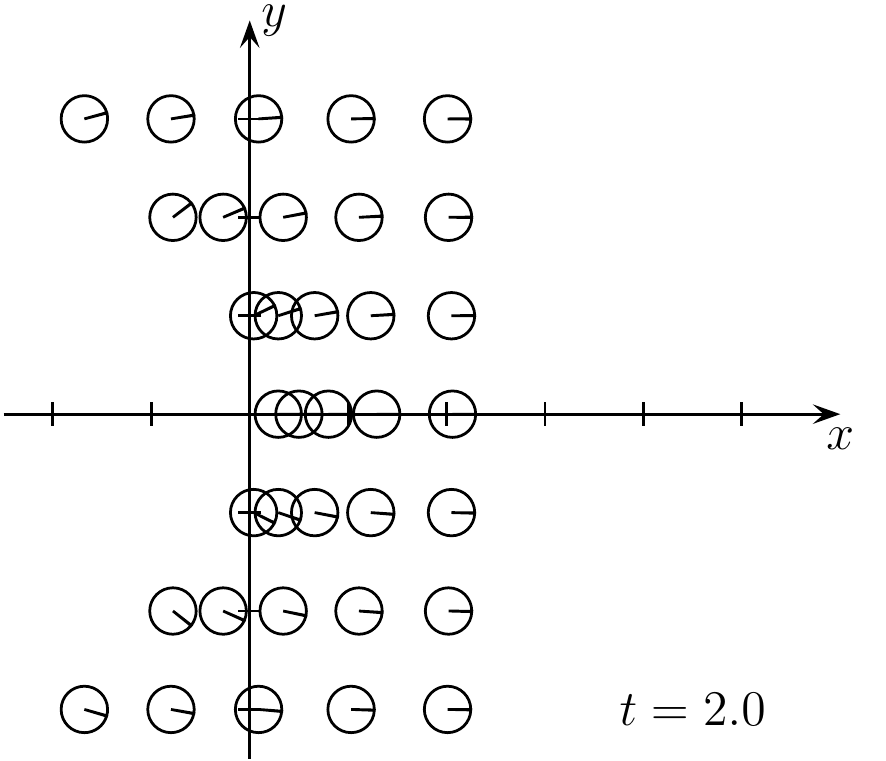}
 \includegraphics[scale=0.45]{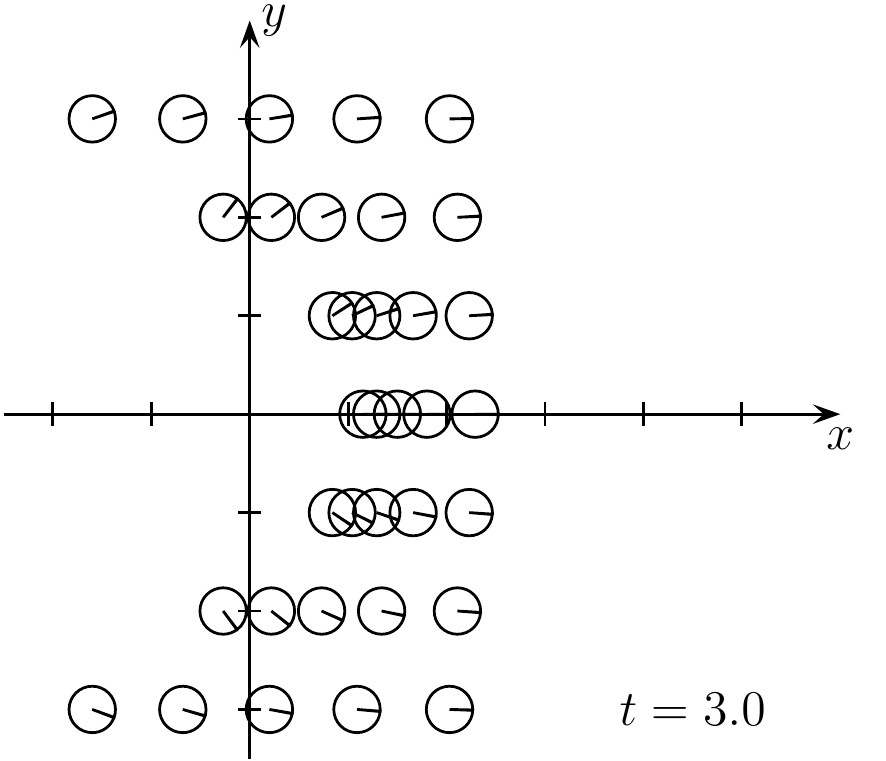}\\[0.3em]
 \includegraphics[scale=0.45]{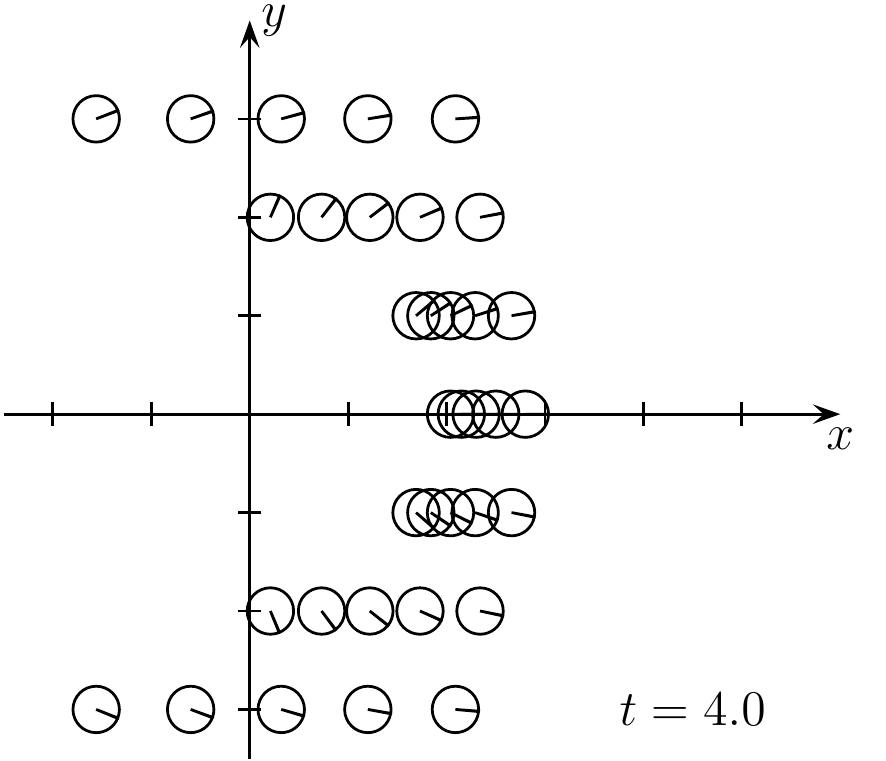}
 \includegraphics[scale=0.45]{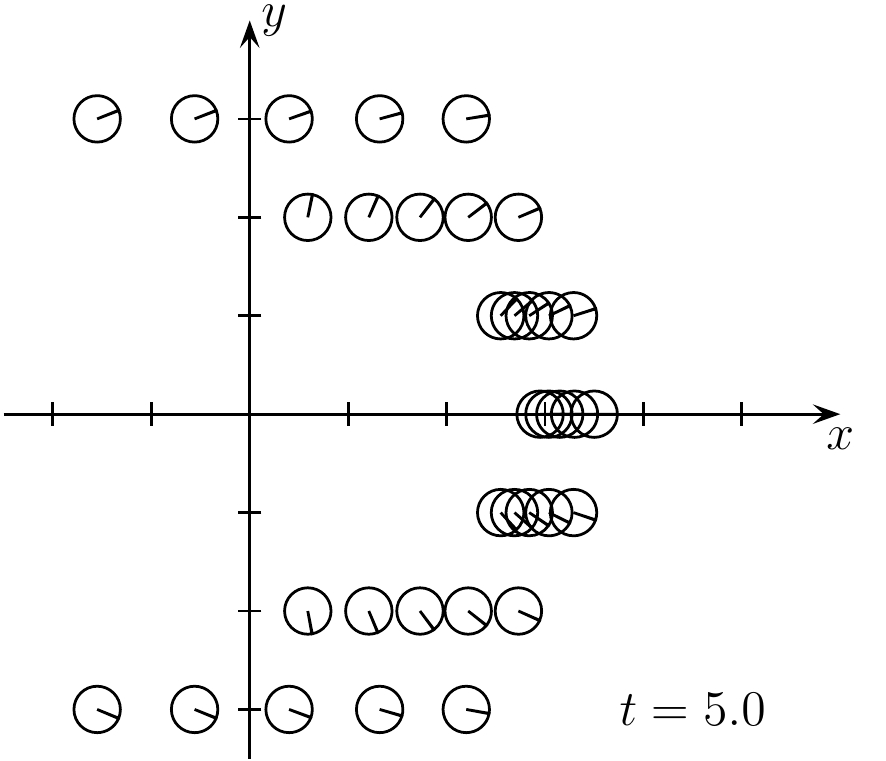}\\[0.3em]
 \includegraphics[scale=0.45]{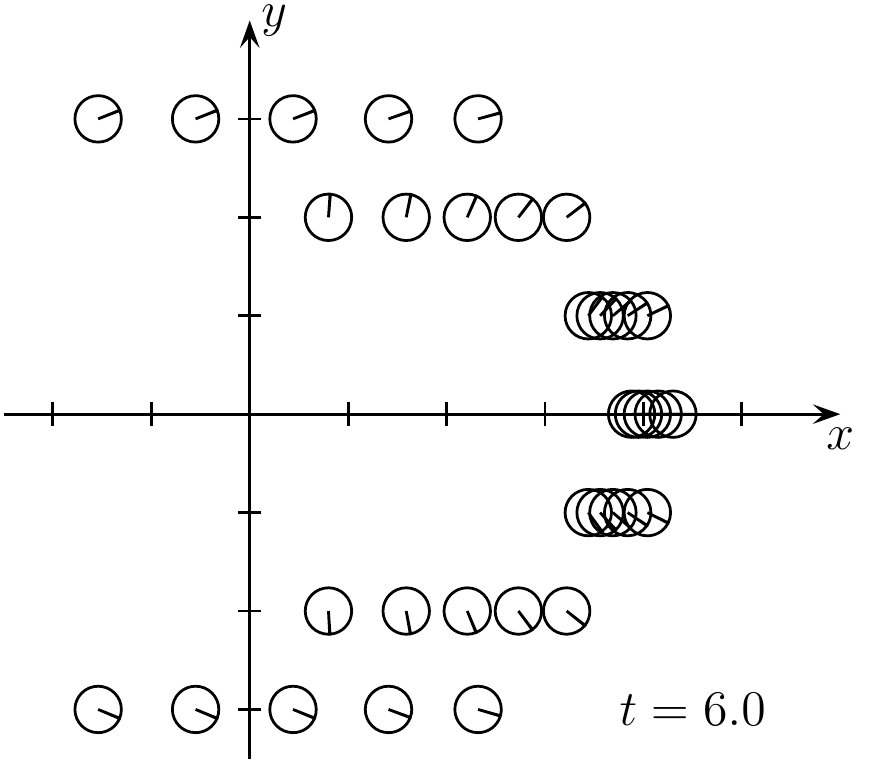}
 \includegraphics[scale=0.45]{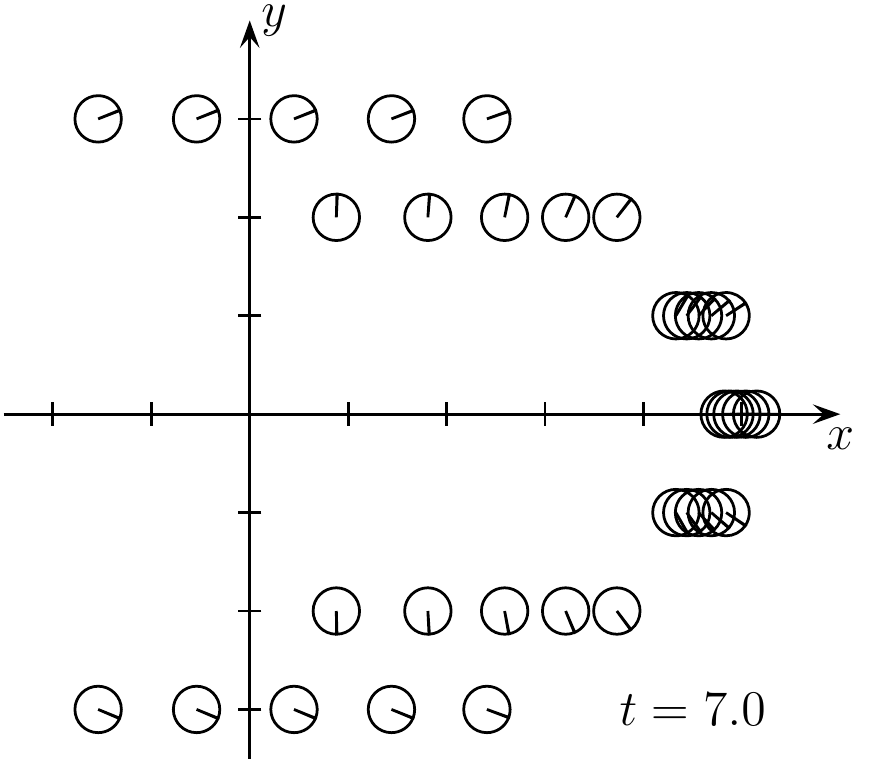}
 \caption{Parallel transport of tiny balls that are initially $(t=-3)$ at rest. The dashes within the balls indicate their orientation. The parameters of the warp metric read $R=2,\sigma=1,v=2c$.}
 \label{fig:partField}
\end{figure}

When the warp bubble approaches the particle field, the particles will be carried along for some time depending on the distance to the center of the warp bubble, see figure \ref{fig:partField}.

Additionally, the particles undergo a geodesic precession away from the $x$-axis while they are in the sphere of influence of the warp bubble. The precession angle is larger the closer the particle is to the rim of the warp bubble. Particles at $y=z=0$ keep their orientation.

% -----------------------------------------------------------------
%                      summary
% -----------------------------------------------------------------
\section{Summary}
The Alcubierre warp spacetime offers a rich playground for studying geodesics in general relativity. For a direct numerical integration of the geodesic equation, the parallel transport, and the Jacobian equation, we have transformed these equations into their non-affinely parametrized form.

The view of an observer comoving with the warp bubble is distorted similarly to the special relativistic aberration. In the direction of motion, however, the distortion is much softer than the special relativistic aberration. Additionally, for velocities larger than the speed of light, there is an event horizon behind the warp bubble. However, there is no black region as in the Schwarzschild spacetime. Nevertheless, the strong redshift as well as the large attenuation let the rear side appear very dark.

The view of a static observer outside the warp bubble heavily depends on his location and observation time. Light rays are strongly deflected and multiple images might appear. 

An interactive special-relativistic flight or warp-drive voyage through the stars is simulated within our Java application. In contrast to the elongated stars shown in science fiction movies, stars would keep their point-like shape.

Particles that start from inside the warp bubble, will be `decelerated' or `accelerated' depending on their initial angle to the warp bubble's direction of motion. Additionally, particles undergo geodesic precession while they are in the sphere of influence of the warp bubble.

%% ---------------------------------------------------------------------------
%%          appendix
%% ---------------------------------------------------------------------------
\appendix

%% ---------------------------------------------------------------------------
%%          Christoffel symbols
%% ---------------------------------------------------------------------------
\section{Christoffel symbols}\label{app:details}
The Christoffel symbols of the Alcubierre metric (\ref{eq:warpMetric}) read
\begin{eqnarray}
 \fl \Gamma^t_{tt} &= \frac{f^2f_xv^3}{c^2}, \quad & \Gamma^y_{tt} = -ff_yv^2, \quad \Gamma^z_{tt} = -ff_zv^2, \quad \Gamma^x_{tt} = \frac{f^3f_xv^4}{c^2}-ff_xv^2-f_tv-f\partial_tv,\\
 \fl \Gamma^t_{tx} &= -\frac{ff_xv^2}{c^2}, & \Gamma^x_{tx} = -\frac{f^2f_xv^3}{c^2}, \quad  \Gamma^y_{tx} = \frac{f_yv}{2}, \quad \Gamma^z_{tx} = \frac{f_zv}{2},\\
 \fl \Gamma^t_{ty} &= -\frac{ff_yv^2}{2c^2}, & \Gamma^x_{ty} = -\frac{f^2f_yv^3+c^2f_yv}{2c^2}, \quad \Gamma^t_{tz} = -\frac{ff_zv^2}{2c^2}, \quad \Gamma^x_{tz} = -\frac{f^2f_zv^3+c^2f_zv}{2c^2},\\
 \fl \Gamma^t_{xx} &= \frac{f_xv}{c^2}, & \Gamma^x_{xx} = \frac{ff_xv^2}{c^2}, \quad \Gamma^t_{xy} = \frac{f_yv}{2c^2}, \quad \Gamma^x_{xy} = \frac{ff_yv^2}{2c^2},\\
 \fl \Gamma^t_{xz} &= \frac{f_zv}{2c^2}, &
 \Gamma^x_{xz} = \frac{ff_zv^2}{2c^2},
\end{eqnarray}
with derivatives
\begin{eqnarray}
 f_t &= \frac{df(r)}{dt} = \frac{-v\left(x-x_s(t)\right) }{r}\frac{df\left(r\right)}{dr},\\
 f_x &= \frac{df(r)}{dx} = \frac{x-x_s(t)}{r}\frac{df\left(r\right)}{dr},\\
 f_y &= \frac{df(r)}{dy} = \frac{y}{r}\frac{df\left(r\right)}{dr},\\
 f_z &= \frac{df(r)}{dz} = \frac{z}{r}\frac{df\left(r\right)}{dr},
\end{eqnarray}
and
\begin{equation}
  \frac{df(r)}{dr} = \frac{\sigma\left[ \sech^2\left( \sigma (r+R)\right) - \sech^2\left( \sigma (r-R)\right) \right]}{2\tanh (\sigma R)}.
\end{equation}
Whether the velocity $v$ of the warp bubble is time-dependent or not has only an influence on the Christoffel symbol $\Gamma_{tt}^x$.

%% ---------------------------------------------------------------------------
%%          
%% ---------------------------------------------------------------------------
\section{Geodesic equation}\label{app:naGeod}
The non-affinely parametrized geodesic equation reads~\cite{nwankwo2010}
\begin{equation}
 \label{appeq:nonAffGeod}
 \frac{d^2x^{\mu}}{d\sigma^2}+\Gamma_{\nu\rho}^{\mu}\frac{dx^{\nu}}{d\sigma}\frac{dx^{\rho}}{d\sigma} = -\frac{1}{\zeta(\sigma)}\frac{d\zeta(\sigma)}{d\sigma}\frac{dx^{\mu}}{d\sigma},
\end{equation}
where $\sigma$ is an arbitrary (non-affine) parameter and $\zeta$ is a function of $\sigma$. The affine parameter $\lambda$ then follows from
\begin{equation}
 \lambda = \int_{\sigma_0}^{\sigma}\frac{a}{\zeta(\sigma)}d\sigma
 \label{appeq:lambda}
\end{equation}
with
\begin{equation}
 \frac{d\sigma}{d\lambda}=\frac{\zeta(\sigma)}{a},\quad \frac{d^2\sigma}{d\lambda^2}=\frac{1}{a}\frac{d\zeta}{d\sigma}\frac{d\sigma}{d\lambda},
\end{equation}
and an arbitrary but constant factor $a$. The relation between the affinely and the non-affinely parametrized geodesic equations can be derived by means of the chain rule
\begin{equation}
 \frac{dx^{\mu}}{d\lambda} = \frac{dx^{\mu}}{d\sigma}\frac{d\sigma}{d\lambda},\quad \frac{d^2x^{\mu}}{d\lambda^2} = \frac{d^2x^{\mu}}{d\sigma^2}\left(\frac{d\sigma}{d\lambda}\right)^2+\frac{dx^{\mu}}{d\sigma}\frac{d^2\sigma}{d\lambda^2}.
\end{equation}

If we replace the non-affine parameter $\sigma$ by the coordinate time $t$, Eq.~(\ref{appeq:nonAffGeod}) with $x^{0}=t$ yields
\begin{equation}
 \label{appeq:zeta}
 \Gamma_{\nu\rho}^0\frac{dx^{\nu}}{dt}\frac{dx^{\rho}}{dt} = -\frac{1}{\zeta(t)}\frac{d\zeta(t)}{dt}.
\end{equation}
Thus, the geodesic equation (\ref{appeq:nonAffGeod}) can be written as
\begin{equation}
 \frac{d^2x^i}{dt^2}+\Gamma_{\nu\rho}^i\frac{dx^{\nu}}{dt}\frac{dx^{\rho}}{dt} = \Gamma_{\nu\rho}^0\frac{dx^{\nu}}{dt}\frac{dx^{\rho}}{dt}\frac{dx^i}{dt},
\end{equation}
which simplifies to 
\begin{eqnarray}
 \nonumber 0 &= \frac{d^2x^i}{dt^2} - \frac{dx^i}{dt}\left(\Gamma_{00}^0+2\Gamma_{0j}^0\frac{dx^j}{dt}+\Gamma_{jk}^0\frac{dx^j}{dt}\frac{dx^k}{dt}\right)\\
  &\quad +\Gamma_{00}^i + 2\Gamma_{0j}^i\frac{dx^j}{dt} + \Gamma_{jk}^i\frac{dx^j}{dt}\frac{dx^k}{dt},
\end{eqnarray}
where $i=1,2,3$. The generalization of the constraint equation (\ref{eq:constrEq}), $g_{\mu\nu}(dx^{\mu}/d\lambda)(dx^{\nu}/d\lambda)=\kappa c^2$ with $\kappa=0$ for light-like and $\kappa=-1$ for time-like geodesics, reads
\begin{equation}
 \left(\frac{dt}{d\lambda}\right)^2\left(g_{00}+2g_{0i}\frac{dx^i}{dt}+g_{ij}\frac{dx^i}{dt}\frac{dx^j}{dt}\right)=\kappa c^2.
\end{equation}
The function $\zeta$ follows from Eq.~(\ref{appeq:zeta})
\begin{equation}
 \zeta(t)=\zeta_0\exp\left[-\int_{t_0}^t\Gamma_{00}^0+\Gamma_{0i}^0\frac{dx^i}{dt'}+\Gamma_{ij}^0\frac{dx^i}{dt'}\frac{dx^j}{dt'}dt'\right].
\end{equation}

%% ---------------------------------------------------------------------------
%%          
%% ---------------------------------------------------------------------------
\section{Euler-Lagrange equations}\label{app:euler}
The Euler-Lagrangian~\cite{rindler} equations for geodesics in the $z=\mbox{const}$ hyperplane with Lagrangian
\begin{equation}
 \mathcal{L} = -c^2\dot{t}^2+\left(\dot{x}-vf\dot{t}\right)^2+\dot{y}^2
\end{equation}
and $v=\mbox{const}$ yield
\begin{eqnarray}
 0 &= \frac{d}{d\lambda}\left[c^2\dot{t}+vf\left(\dot{x}-vf\dot{t}\right)\right]-\left(\dot{x}-vf\dot{t}\right)\dot{t}vf_t,\\
 0 &= \frac{d}{d\lambda}\left[\dot{x}-vf\dot{t}\right]+\left(\dot{x}-vf\dot{t}\right)v\dot{t}f_x,\\
 0 &= \ddot{y} + \left(\dot{x}-vf\dot{t}\right)v\dot{t}f_y,
\label{appeq:el}
\end{eqnarray}
where a dot means differentiation with respect to the affine parameter $\lambda$. It is obvious that these equations are automatically fulfilled for
\begin{equation}
 \dot{t}=k_1,\quad \dot{x}=vfk_1,\quad \dot{y}=k_2,
 \label{appeq:elInit}
\end{equation}
with constants of motion $k_1$ and $k_2$. These initial values correspond to a null or time-like geodesic that starts perpendicular to the direction of motion with respect to the comoving reference frame of the center of the bubble. Because $y$ grows linearly, the shape function $f$ tends to zero and, thus, $x$ is limited. Hence, such a geodesic approaches a line orthogonal to the $x$-axis.

In case of a null geodesic, the initial direction reads $\mathbf{k}=\omega\left(\pm\mathbf{e}_{(0)}+\mathbf{e}_{(2)}\right)$, which yields $k_1=\pm \omega/c$ and $k_2=\omega$. Thus, there is no frequency shift.

A time-like geodesic with initial local velocity $v_{\mbox{part}}$ and four-velocity $\mathbf{u}=c\gamma(\mathbf{e}_{(0)}+\beta\mathbf{e}_{(2)})$, where $\beta=v_{\mbox{part}}/c$ and $\gamma=1/\sqrt{1-\beta^2}$, has constants of motion $k_1=\gamma$ and $k_2=\gamma\beta c$.

For null geodesics that are restricted to the $x$-axis, Eq.~(\ref{appeq:el}) simplifies to
\begin{equation}
 \frac{d}{d\lambda}\dot{t}+vf_x\dot{t}^2=0\quad\mbox{and}\quad \frac{d}{d\lambda}\dot{x}+v^2f_x\dot{t}^2=0,
\end{equation}
where we made use of $\mathcal{L}=0$, $\dot{y}=0$, and $f_t=-vf_x$. Thus, a future-directed initial direction $\mathbf{k}=\omega\left(\mathbf{e}_{(0)}\pm\mathbf{e}_{(1)}\right)$ leads to $\dot{x}-v\dot{t}=k_3=\omega\left[\frac{v}{c}(f-1)\pm 1\right]$, which is a constant of motion, and we obtain
\begin{equation}
 \dot{t}=\frac{\pm k_3}{c\mp v(1-f)},\quad \dot{x}=k_3\frac{c\pm vf}{c\mp v(1-f)}.
\end{equation}

%% ---------------------------------------------------------------------------
%%          
%% ---------------------------------------------------------------------------
\section{Parallel transport of Sachs basis and Jacobi equation}\label{app:sachsjac}
The parallel transport equation for the Sachs basis vector $\mathbf{s}=s^{\mu}\partial_{\mu}$ can be cast into the form
\begin{equation}
 0 = \frac{ds^{\mu}}{dt} + \left(\Gamma_{0\nu}^{\mu}+\Gamma_{i\nu}^{\mu}\frac{dx^i}{dt}\right)s^{\nu}.
\end{equation}
The Jacobian equation for the Jacobian field $\mathbf{Y}=Y^{\mu}\partial_{\mu}$ reads
\begin{eqnarray}
\label{appeq:jacobian} 
\nonumber 0 &= \frac{d^2Y^{\mu}}{dt^2} - \frac{dY^{\mu}}{dt}\left(\Gamma_{00}^0+2\Gamma_{0j}^0\frac{dx^j}{dt}+\Gamma_{jk}^0\frac{dx^j}{dt}\frac{dx^k}{dt}\right)\\ 
  &\quad + 2\left(\Gamma_{o\nu}^{\mu} + \Gamma_{i\nu}^{\mu}\frac{dx^i}{dt}\right)\frac{dY^{\nu}}{dt}\\
 \nonumber &\quad + \left(\Gamma_{00,\nu}^{\mu} + 2\Gamma_{0i,\nu}^{\mu}\frac{dx^i}{dt} + \Gamma_{ij,\nu}^{\mu}\frac{dx^i}{dt}\frac{dx^j}{dt}\right)Y^{\nu}.
\end{eqnarray}
The initial Sachs basis vectors are perpendicular to the initial light direction $\mathbf{k}=-\mathbf{e}_{(0)}+\cos\xi\mathbf{e}_{(1)}+\sin\xi\mathbf{e}_{(2)}$, thus 
\begin{equation}
  \mathbf{s}_1=-\sin\xi\mathbf{e}_{(1)}+\cos\xi\mathbf{e}_{(2)}\quad\mbox{and}\quad \mathbf{s}_2=\mathbf{e}_{(3)}.
  \label{eq:sachsVec}
\end{equation}
The initial values for the two Jacobian fields read $Y^{\mu}_{1,2}\big|_{t=0}=0$ and $dY_{1,2}^{\mu}/{dt}\big|_{t=0}=s_{1,2}^{\mu}$.

\section{Numerical integration}
The integration of the non-affinely parametrized geodesic equation avoids the problem of the inappropriate affine parameter. However, we still need the affine parameter for the calculation of the magnification factor. From Eqs. (\ref{appeq:lambda}) and (\ref{appeq:zeta}) we can expand the set of ordinary differential equations for $\lambda$ to 
\begin{eqnarray}
   \frac{d\lambda}{dt} &= \frac{a}{\zeta},\\
   \label{appeq:lzeta}\frac{d\zeta}{dt} &= \left(-\Gamma_{00}^0 - 2\Gamma_{0i}^0\frac{dx^i}{dt} - \Gamma_{ij}^0\frac{dx^i}{dt}\frac{dx^j}{dt}\right)\zeta.
\end{eqnarray}
Since $\zeta$ grows exponentially, which results in numerical problems, we substitute $\zeta=\exp(\psi)$ in (\ref{appeq:lzeta}) and obtain
\begin{equation}
 \frac{d\psi}{dt} = -\Gamma_{00}^0 - 2\Gamma_{0i}^0\frac{dx^i}{dt} - \Gamma_{ij}^0\frac{dx^i}{dt}\frac{dx^j}{dt}.
\end{equation}
Here, we set $a=1$.

Numerical problems due to exponential grow of the Sachs vectors and the Jacobi functions make the following substitutions necessary:
\begin{equation}
 s^{\mu} = \sinh(p^{\mu}),\quad Y^{\mu} = \sinh(u^{\mu}).
\end{equation}
The parallel transport of the Sachs vectors thus reads
\begin{equation}
 \frac{dp^{\mu}}{dt} = -\left(\Gamma_{0\sigma}^{\mu}+\Gamma_{i\sigma}^{\mu}\frac{dx^i}{dt}\right)\frac{\sinh(p^{\sigma})}{\cosh(p^{\mu})}.
\end{equation}
Note that there is no summation over the index $\mu$ in the right hand side of this equation.

For the integration of the Jacobian equation (\ref{appeq:jacobian}), we obtain
\begin{eqnarray}
   \frac{du^{\mu}}{dt} &= w^{\mu},\\
   \frac{dw^{\mu}}{dt} &= -\tanh(u^{\mu})\left(w^{\mu}\right)^2\\
 \nonumber   &\quad + \left(\Gamma_{00}^0+2\Gamma_{0j}^0\frac{dx^j}{dt}+\Gamma_{jk}^0\frac{dx^j}{dt}\frac{dx^k}{dt}\right)w^{\mu}\\
 \nonumber   &\quad - 2\left(\Gamma_{o\nu}^{\mu} + \Gamma_{i\nu}^{\mu}\frac{dx^i}{dt}\right)\frac{\cosh(u^{\nu})}{\cosh(u^{\mu})}w^{\mu}\\
 \nonumber   &\quad - \left(\Gamma_{00,\nu}^{\mu} + 2\Gamma_{0i,\nu}^{\mu}\frac{dx^i}{dt} + \Gamma_{ij,\nu}^{\mu}\frac{dx^i}{dt}\frac{dx^j}{dt}\right)\frac{\sinh(u^{\nu})}{\cosh(u^{\mu})}.
\end{eqnarray}
As before, there is no summation over the index $\mu$.

To apply numerical libraries like, for example, the Gnu Scientific Library~\cite{gsl} or the numerical recipes~\cite{numrec}, we have to map the above equations to a one-dimensional array as follows:
\begin{eqnarray}
\label{eq:numSystem}
  \mathtt{y}[n] &= x^{n+1}, &\quad \mathtt{y}[n+3] = \frac{dx^{n+1}}{dt},\\
  \mathtt{y}[m+6] &= p_1^m, & \mathtt{y}[m+10] = p_2^m,\\
  \mathtt{y}[m+14] &= Y_1^m, & \mathtt{y}[m+18] = \frac{dY_1^m}{dt},\\
  \mathtt{y}[m+22] &= Y_1^m, & \mathtt{y}[m+26] = \frac{dY_1^m}{dt},\\
  \mathtt{y}[30] &= \psi, & \mathtt{y}[31] = \lambda.
\end{eqnarray}
Here, $n=0,1,2$ and $m=0,1,2,3$. A light ray with initial direction $\mathbf{k}=-\mathbf{e}_{(0)}+\cos\xi\mathbf{e}_{(1)}+\sin\xi\mathbf{e}_{(2)}$ yields $\dot{t}=-1/c$, $\dot{x}=-vf/c+\cos\xi$, and $\dot{y}=\sin\xi$. Thus $\mathtt{y}[3]=\dot{x}/\dot{t}=vf-c\cos\xi$, $\mathtt{y}[4]=-c\sin\xi$, and $\mathtt{y}[5]=0$. The corresponding Sachs vectors (\ref{eq:sachsVec}) give $\mathtt{y}[6]=\mathtt{y}[9]=0$, $\mathtt{y}[7]=\arsinh(-\sin\xi)$, $\mathtt{y}[8]=\arsinh(\cos\xi)$, $\mathtt{y}[10]=\mathtt{y}[11]=\mathtt{y}[12]=0$, and $\mathtt{y}[13]=\arsinh(-1)$. The initial values for the Jacobi vector fields are obvious. For the integration of the affine parameter $\lambda$ with $\lambda(0)=0$, we have $\zeta(0)=1$ and $\psi(0)=0$. Hence, $\mathtt{y}[30]=\mathtt{y}[31]=0$.

For the numerical integration of time-like geodesics, we can reduce the system (\ref{eq:numSystem}) to only eight equations. The array elements $\mathtt{y}[]$ can then be deduced from the following equations. Particles from the bridge have initial directions
\begin{equation}
 \frac{dt}{d\lambda}=\gamma,\quad \frac{dx}{dt}=vf+c\beta\cos\xi,\quad \frac{dy}{dt}=c\beta\sin\xi,
\end{equation}
whereas particles injected from a static outside observer are described by
\begin{eqnarray}
  \frac{dt}{d\lambda} &= \frac{c\gamma}{\sqrt{c^2-v^2f^2}}\left(1-\beta vf\cos\xi\right),\\
  \frac{dx}{dt} &= \gamma\beta\sqrt{c^2-v^2f^2}\cos\xi,\quad \frac{dy}{dt} = c\gamma\beta\sin\xi.
\end{eqnarray}
In both cases, we have $\psi(0)=\ln\gamma$.

%% ------------------------------------------------------------------------
%%                               acknowledgments
%% ------------------------------------------------------------------------
\ack
This work was partially funded by Deutsche Forschungsgemeinschaft (DFG) as part of the Collaborative Research Centre SFB 716 and the DFG project ``Astrographik''.

% -----------------------------------------------------------------
%                           thebibliography
% -----------------------------------------------------------------

\end{document}